\documentclass[prb,a4paper,superscriptaddress,twocolumn,showkeys,amsmath,floatfix]{revtex4-2}
\usepackage{graphicx}
\usepackage{amssymb}
\usepackage{float}
\usepackage{xurl}
\usepackage{booktabs}
\usepackage{color}
\usepackage{hyperref}
\hypersetup{
    colorlinks=true,
    linkcolor=blue,
    urlcolor=blue,
    citecolor=blue
    }
\begin{document}
\title{Dollar-Yuan Battle in the World Trade Network}
\author{C\'{e}lestin Coquid\'{e}}
\author{Jos\'e Lages} 
\email[To whom correspondence should be addressed. E-mail: ]{jose.lages@univ-fcomte.fr}
\affiliation{Institut UTINAM, OSU THETA, Universit\'e Bourgogne Franche-Comt\'e, CNRS, Besançon, France}
\author{Dima L. Shepelyansky}
\affiliation{\mbox{Laboratoire de Physique Th\'eorique,
Universit\'e de Toulouse, CNRS, UPS, 31062 Toulouse, France}}

\date{This manuscript was compiled on \today}
\begin{abstract}%	Please provide an abstract of no more than 250 words in a single paragraph. Abstracts should explain to the general reader the major contributions of the article. References in the abstract must be cited in full within the abstract itself and cited in the text.
\textbf{Abstract} -- From the Bretton Woods agreement in 1944 till the
	present day, the US dollar has been the dominant
	currency in the world trade.
	However, the rise of the Chinese economy 
	led recently to the emergence of
	trade transactions in Chinese yuan.
	Here, we analyze mathematically how the structure of the international trade flows would favor a country to trade whether in US dollar or in Chinese yuan. The computation of the trade currency preference is based on the world trade network built from the 2010-2020 UN Comtrade data.
	The preference of a country to trade in US dollar or Chinese yuan is determined by two multiplicative factors: the relative weight of trade volume exchanged by the country with its direct trade partners, and the relative weight of its trade partners in the global international trade.
	The performed analysis, based on
	Ising spin interactions on the world trade network,
	shows that, from 2010 to present,
	a transition took place, and the
	majority of the world countries
	would have now a preference
	to trade in Chinese yuan if one only consider the world trade network structure.
\end{abstract}
\maketitle
\section{Introduction}
In 1944, the Bretton Woods agreements established a system of payments based on the US dollar (USD). 
The USD effectively became the world currency, i.e.,
the standard to which every other currency was pegged \cite{wiki1}.
Till now, the USD remained the dominant world trade currency and, as an example, the
United Nations (UN) reports the world trade transactions between countries in USD \cite{comtrade}.
However, the possible end of the dollar dominance is increasingly discussed as the Chinese yuan (CNY) is gradually becoming credible as a reserve currency \cite{FP,BS,GT,balance,bloomberg,liu22,nikkei}.
Moreover, recently important trade transactions were considered to be
realized in CNY instead of USD like oil sales from Saudi Arabia to China \cite{wallstrj}.
Thus, the growth of the Chinese economy \cite{chinaecon} opens up the possibility for a country to prefer the CNY to the USD for their trade exchanges.

The World
Trade Organization (WTO) Statistical Review \cite{WTOSR22}
demonstrates the vital importance of international
trade for the development and
progress of the world countries. As the world economy deeply depends
on the world trade \cite{krugman11}, the detailed analysis of
the trade flows with a possible switch of the trade currency leads to significant effects on the world economy which can be used by policy makers and other stakeholders.
Although the choice of the invoicing currency is generally taken at the importing/exporting firms level, for the sake of simplicity, our study is based on a model where each country, as a whole, takes the decision to trade in a given currency.

In the present paper,
we model the trade currency
preference of a country, ie, here whether a country prefers to trade in USD or in CNY, as a binary variable with the properties of a spin in an Ising model. In absence of any political considerations or extra economical factors, here we consider that a country will prefer to trade in one currency rather than in the other one if the structure of the international trade indeed favors the former. We study how the preference for a trade currency spreads from country to country with the help of a model derived from opinion formation models \cite{galam08,castellano09,opinion,schmittmann10}.
This currency preference model is applied on the world trade network (WTN) built from the 2010-to-2020 UN Comtrade data \cite{comtrade}. 
The constructed WTN describes trade relations between countries in terms of entropy flows.
Our analysis, based purely on trade relations
without taking account of external political factors or other considerations, clearly shows that, during this decade, a USD-to-CNY transition took place which imply that the structure of the WTN would now favor trades in CNY rather than in USD.

\section{Data sets and model description}
The network analysis of various types of flows finds many applications in various areas of science
(see e.g. \cite{dorogovtsev,meyer,rmp2015}).
More specifically, the network analysis of the world trade transactions
has been performed e.g. in \cite{dorogovtsev,serrano07,fagiolo09,he10,fagiolo10,barigozzi10,chakraborty18,debenedictis11,wtn1,wtn3,wtncrisis}.

The WTN is constructed from
the UN Comtrade database \cite{comtrade} which provides the money matrix $\mathcal{M}$ encoding the transactions of all the commodities between all the countries. The money matrix element $M_{c'c}$ gives the total amount of commodities, expressed in USD of a given year, exported from the country $c$ to the country $c'$. The UN Comtrade database concerns 194 countries
for the period 2010-2020. The Markov chain of trade transactions
is characterized by two WTN trade matrices $\mathcal{S}$ and $\mathcal{S^*}$ whose elements are
$S_{c'c}=M_{c'c}/M^*_c$
and 
$S^*_{c'c}=M_{cc'}/M_c$. Here, $M^*_c=\sum_{c'}M_{c'c}$  ($M_c=\sum_{c'}M_{cc'}$) gives the total amount of commodities
exported from (imported to) the country $c$ to (from) the rest of the world.
Consequently, the matrix element $S_{c'c}$ ($S^*_{c'c}$) gives the relative weight of the imports to (exports from) the country $c'$ from (to) the country $c$.
These matrix elements can be considered as link weights
of a directed network
(see e.g. \cite{dorogovtsev}).
By construction, the sum of each column of the matrices $\mathcal{S}$ and $\mathcal{S^*}$ is equal to 1, i.e., $\sum_{c'}S_{c'c}=1$ and $\sum_{c'}S^*_{c'c}=1$, ensuring the fact that the stochastic processes of the trade transactions, both in export and import directions, belong to the class of Markov chains. Let us define, for a given year, the total world trade volume by
$M=\sum_c M_c = \sum_c M^*_c$.
Then, a given country $c$ can be characterized
by its import trade probability $P_c = M_c/M$ and its export trade probability $P^*_c = M^*_c/M$.
These probabilities characterize the global capability of a given country to import and to export, respectively.
As an example, in 2019, the top 5 countries according to the import trade probability $P_c$ are: 1. USA, 2. China, 3. Germany, 4. Japan, 5. UK, and the top 5 Countries according to the export trade probability $P^*_c$ are:
1. China, 2. USA, 3. Germany, 4. Japan, 5. France.

In order to determine the trade currency preference (TCP) of the country $c$, i.e., whether the country $c$ would prefer to trade in USD or in CNY with the other world countries, we assign to the country $c$ an Ising spin $\sigma_c=\pm1$. The value $\sigma_c=-1$ ($+1$) indicates that the country $c$ prefers to trade in USD (CNY). Thus, we obtain a network of interacting Ising spins, each one attached to a country. We define the interaction energy of the country (spin) $c$ as
\begin{equation}
	E_c = \frac12\sum_{c' \neq c} \sigma_{c'} \left(S^{\phantom{*}}_{c'c} + S^*_{c'c}\right) \left(P^{\phantom{*}}_{c'} +P^*_{c'}\right)
	\label{eq1}
\end{equation}
which is the sum of all the energies of interaction of the country (spin) $c$ with its direct trade partners (spins) $c'$.
The country $c$ TCP possibly flips form one currency to the other according to the sign of the computed $E_c$: If $E_{c} < 0$ then 
	the country $c$ spin becomes $\sigma_c=-1$ 
	(preference to trade in USD)
	and
	if $E_c >0$ then
	the country $c$ spin becomes $\sigma_c=+1$
	(preference to trade in CNY).
From the interaction energy $E_c$ expression (\ref{eq1}), the formation of the country $c$ TCP is sensitive to the TCPs of the direct commercial partners, through the term $\sigma_{c'}$, weighted by two important factors:
\begin{itemize}
\item the ``$S^{\phantom{*}}_{c'c} + S^*_{c'c}$'' term which encodes the relative strength of the import-export flows between the country $c$ and a direct trade partner $c'$.
As an example, in 2010,
0.4\% of the total volume of exports from Russia was imported by Brazil ($S_{\rm BR,RU}=0.004$) and
1.9\% of the total volume of exports from Brazil was imported by Russia ($S_{\rm RU,BR}=0.019$). Also,
1.7\% of total volume of imports to Russia was exported from Brazil ($S^*_{\rm BR,RU}=0.017$) and
1.0\% of total volume of imports to Brazil was exported from Russia ($S^*_{\rm RU,BR}=0.017$).
\item the ``$P^{\phantom{*}}_{c'} +P^*_{c'}$'' term which encodes the global trade capability of the partner $c'$ in the WTN. As an example, in 2010, the total imports to Brazil and to Russia represent 1.3\% ($P_{\rm BR}=0.013$) and 1.7\% ($P_{\rm RU}=0.017$) of the total world trade volume $M$,
and the total exports from Brazil and from Russia represent 1.5\% ($P^*_{\rm BR}= 0.015$) and 3.1\% ($P^*_{\rm RU}=0.031$) of the total trade volume $M$.
\end{itemize}

Consequently, as an example, the direct contribution of China to the possible change of TCP of Russia comes from the $(S^{\phantom{*}}_{\rm CN,RU} + S^*_{\rm CN,RU})(P^{\phantom{*}}_{\rm CN} +P^*_{\rm CN})$ term in (\ref{eq1}) which changes from $0.01$ in 2010
to $0.014$ in 2919 indicating
a significant increase of trade
exchange between Russia and China.
This 40\% increase is partly the reason why in 2010 the Russia's TCP still depends on the initial distribution of the TCPs over all the countries (in 2010, Russia belongs to the hereafter defined swing group, see the Section~\ref{sec:result}) and in 2019 the Russia's TCP is always CNY independently on the initial TCPs distribution (in 2019, Russia belongs to the hereafter defined CNY group, see the Section~\ref{sec:result}). Of course, the Russia's TCP also depends on the other $(S^{\phantom{*}}_{c'c} + S^*_{c'c})(P^{\phantom{*}}_{c'} +P^*_{c'})$ terms in (\ref{eq1}) associated to the other than China countries $c'$.

In this model, we keep USA and China
always trading in USD and CNY respectively.
We start with an initial fraction $f_i$ of randomly chosen countries which prefer to trade in USD, we assign a $-1$ value to their spins. Consequently, the complementary fraction $1-f_i$ of countries prefer to trade in CNY, we assign a $+1$ value to their spins.
A first Monte Carlo shake allows to determine the energy $E_{c_1}$ for a randomly picked spin $\sigma_{c_1}$. The spin $\sigma_{c_1}$ flips or not according to the sign of the newly computed $E_{c_1}$, and consequently the TCP of the country $c_1$ either stay the same or possibly changes from one currency to the other (i.e., from USD to CNY or from CNY to USD).
Then, with the obtained new configuration of spins, a second shake is performed for another randomly chosen spin $\sigma_{c_2}$ $(c_2\neq c_1)$, and so on. After 192 shakes, the trade currency preference for each country is determined
(USA and China spins are always kept fixed).
The ensemble of these shakes forms the first time step $\tau=1$.
We observe that after at most five consecutive time steps $(\tau>5)$ the system converges to a fixed steady-state
configuration of spins which stays unchanged
for higher values of $\tau$.
This procedure is applied to $N_r=10^4$ initial random spin configurations
with a fixed fraction $f_i$ of countries with initially a TCP for USD. We follow the evolution with $\tau$ of the fraction $f(\tau)$ of countries preferring to trade in USD till the
steady-state $f_f$ at $\tau=10$ is obtained. The complementary fraction $1-f_f$ gives then the fraction of countries preferring to trade in CNY once the steady-state is reached.
An example of time evolution $f(\tau)$ is shown in Fig.~\ref{figS1}
of the Supplementary Information.

The above described TCP formation model applied on the WTN is similar to models of opinion formation on social networks used to study voting systems, strike phenomena, coalitions formation (see eg \cite{galam08} and \cite{castellano09} for reviews), or opinion propagation in the WWW or Twitter \cite{opinion}. Our model belongs to the class of Ferromagnetic Ising spins \cite{krapivsky10} models previously used to described opinion dynamics \cite{galam08,castellano09,opinion,schmittmann10} on regular and complex networks.
Usually an Ising spin, with a randomly chosen value $-1$ or $+1$, is assigned to each node. The possible flip of a spin, ie the possible change of opinion of a node of the network, taken at random, affects possibly the opinion of its direct neighbors. As in Ferromagnetic Ising spins models at low temperature, after multiple avalanches of spin flips, 
a long range order is established all over the network. As a consequence, one or more giant homogeneous opinion clusters are formed inside which the nodes behave collectively. Possibly, by percolation, one of the opinion clusters dominates and spans the most of the network. A review can be found in, eg, Chapter 8 of \cite{castellano09}.

In the frame of Ferromagnetic Ising spins models
it is very natural that a spin oriented up
and surrounded by
spins oriented down will be also flipped down \cite{galam08,castellano09,opinion,schmittmann10}.
This physical process is directly implemented in the opinion
formation models \cite{galam08,castellano09,opinion,schmittmann10}
and we also use this flip rule in our mathematical model
based on the equation (1). Indeed, if a country has
mainly trade partners preferring to trade in CNY
it naturally would prefer also to trade in CNY
instead of USD. Such an approach to analysis of opinion formation
in social networks
is broadly used in the literature \cite{galam08,castellano09,opinion,schmittmann10}
and we simply extend and apply it here to
the WTN.

\begin{figure}[t]
	\begin{center}
		\includegraphics[width=\columnwidth]{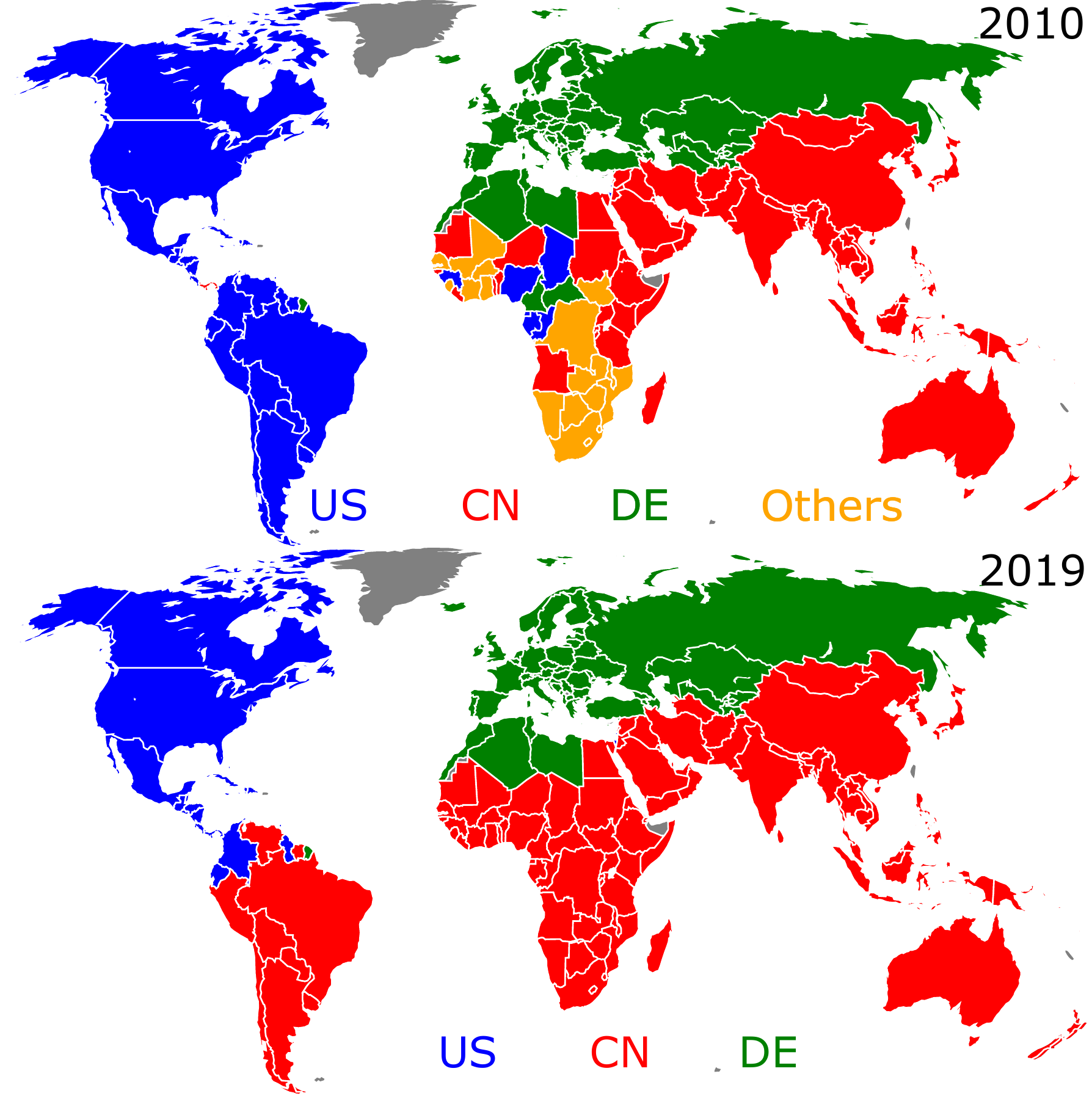}
	\end{center}
	\vglue -0.3cm
	\caption{\label{fig1}Geographical distribution of clusters in the WTN for the years 2010 (left panel) and 2019 (right panel) obtained using the Louvain modularity method \cite{blondel08} with Dugu\'e's algorithm \cite{dugue15}.
		Each cluster is labelled using the ISO2 code of the country with the best import and export trade probabilities $P_c$ and $P^*_c$.
		The leaders of the clusters are the USA (blue), China (red), and Germany (green).
		The cluster Others (gold) gathers other small size clusters.}
\end{figure}
\section{Results}
\label{sec:result}
As a preliminary step, let us characterize, for the years 2010 and 2019, the WTN using the Louvain method for cluster detection \cite{blondel08,dugue15}.
The results presented in Fig.~\ref{fig1} show
the existence of 3 main clusters formed around the USA, China and Germany.
We observe that from 2010 to 2019 the size of the cluster
around China extends significantly. Indeed, from 2010 to 2019, the US-cluster loses
almost all of the South American countries to the benefit of the CN-cluster, and a dominant part of Africa enters the CN-cluster.
Meanwhile, the cluster formed around Germany
remains practically unchanged including the
EU countries, the countries of the former Soviet Union, and most of the Maghreb. Examples of WTN clustering
for other years of the considered decade are shown in Fig.~\ref{figS2} of the Supplementary Information.

\begin{figure}[t]
	\begin{center}
		\includegraphics[width=\columnwidth]{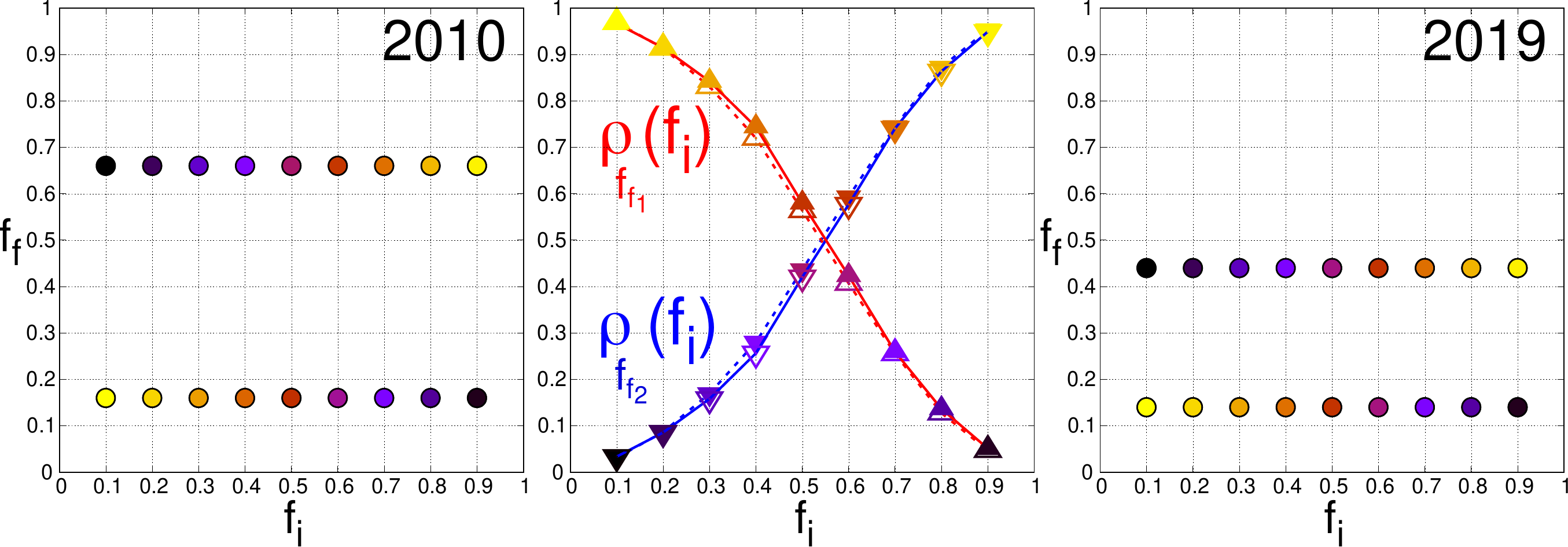}
	\end{center}
	\vglue -0.3cm
	\caption{\label{fig2}Final fraction $f_f$ of countries preferring to trade in USD as a function of the initial fraction $f_i$ for the years 2010 (left panel) and 2019 (right panel). Left and right panels: for any initial fraction $f_i$, the Monte Carlo procedure converges toward one of two final fractions $f_f$. These two final fractions are $f_{f_1} = 0.16$ and $f_{f_2}=0.66$ in 2010 (left panel)
		and $f_{f_1}=0.14$ and $f_{f_2}=0.44$ in 2019 (right panel). The color of a point ($f_i$,$f_f$) indicates the portion $\rho_{f_f}(f_i)$ of the $N_r=10^4$ initial configurations, with a corresponding initial fraction $f_i$, which attain the final state with the corresponding final fraction $f_f$. The color ranges from black for $\rho_{f_f}(f_i)=0$ (all the countries preferring to trade in CNY rather than in USD) to bright yellow for $\rho_{f_f}(f_i)=1$ (all the countries preferring to trade in USD rather than in CNY). Middle panel: Portion $\rho_{f_f}(f_i)$ of the $N_r=10^4$ initial configurations, the fraction $f_i$ of which initially prefers to trade in USD, which attains the final state with the final fraction $f_f$. The red (blue) curve and the up (down) triangles correspond to the lowest (highest) value $f_{f_1}$ ($f_{f_2}$) of the two final fractions $f_f$. The full (empty) symbols correspond to the year 2019 (2010).
	}
\end{figure}

However, this preliminary cluster analysis of the WTN, based on the maximization of the modularity \cite{blondel08,dugue15}, does not determine the trade preference of the countries either for USD or CNY.
The Monte Carlo procedure, described in the previous section, allows to obtain the final fraction $f_f$ of world countries which prefers to trade in USD, and conversely the final fraction $1-f_f$ of world countries which prefers to trade in CNY. Fig.~\ref{fig2} shows the final fraction $f_f$ as a function of the fraction $f_i$ of countries which initially prefer to trade in USD. For each value of $f_i$, we randomly picked $N_r=10^4$ different configurations of spins.
We observe that for any initial fraction $f_i$ belonging to the interval $[0,1]$, only two final fractions $f_{f_1}$ and $f_{f_2}$ can be reached. However, the probability that a given spin configuration reach one or the other final fraction values depends on the initial distribution of the TCPs over the countries. Let us take $f_{f_1}<f_{f_2}$.
Quite naturally, higher (lower) is the initial fraction $f_i$, higher is the probability to obtain the highest (lowest) final value $f_{f_2}$ ($f_{f_1}$).
The middle panel of Fig.~\ref{fig2} gives, for the years 2010 and 2019, the probabilities $\rho_{f_{f_1}}(f_i)$ and $\rho_{f_{f_2}}(f_i)$ to obtain the final fraction $f_{f_1}$ and $f_{f_2}$ as a function of the initial fraction $f_i$.
%For the sake of completeness, Fig.~S3 in Supplementary Information gives, for the years 2010 and 2019, the probability $\rho(f_i)$ to obtain the highest fraction $f_f$ as a function of the initial fraction $f_i$.
In 2010, see left panel of Fig.~\ref{fig2}, each of the final fractions corresponded to a majority of countries with either a USD preference ($f_{f_2}=0.66$) or a CNY preference ($f_{f_1}=0.16$). This no longer the case in 2019, see right panel of Fig.~\ref{fig2}, for which the two final fractions $f_{f_1}=0.14$ and $f_{f_2}=0.44$ are below $0.5$ and give both a CNY preference for the majority of the world countries. In one decade and according to the sole structure of the WTN, we pass from a bipolar USD-CNY trade currency preference to a global domination of the CNY.

\begin{figure}[t]
	\begin{center}
		\includegraphics[width=\columnwidth]{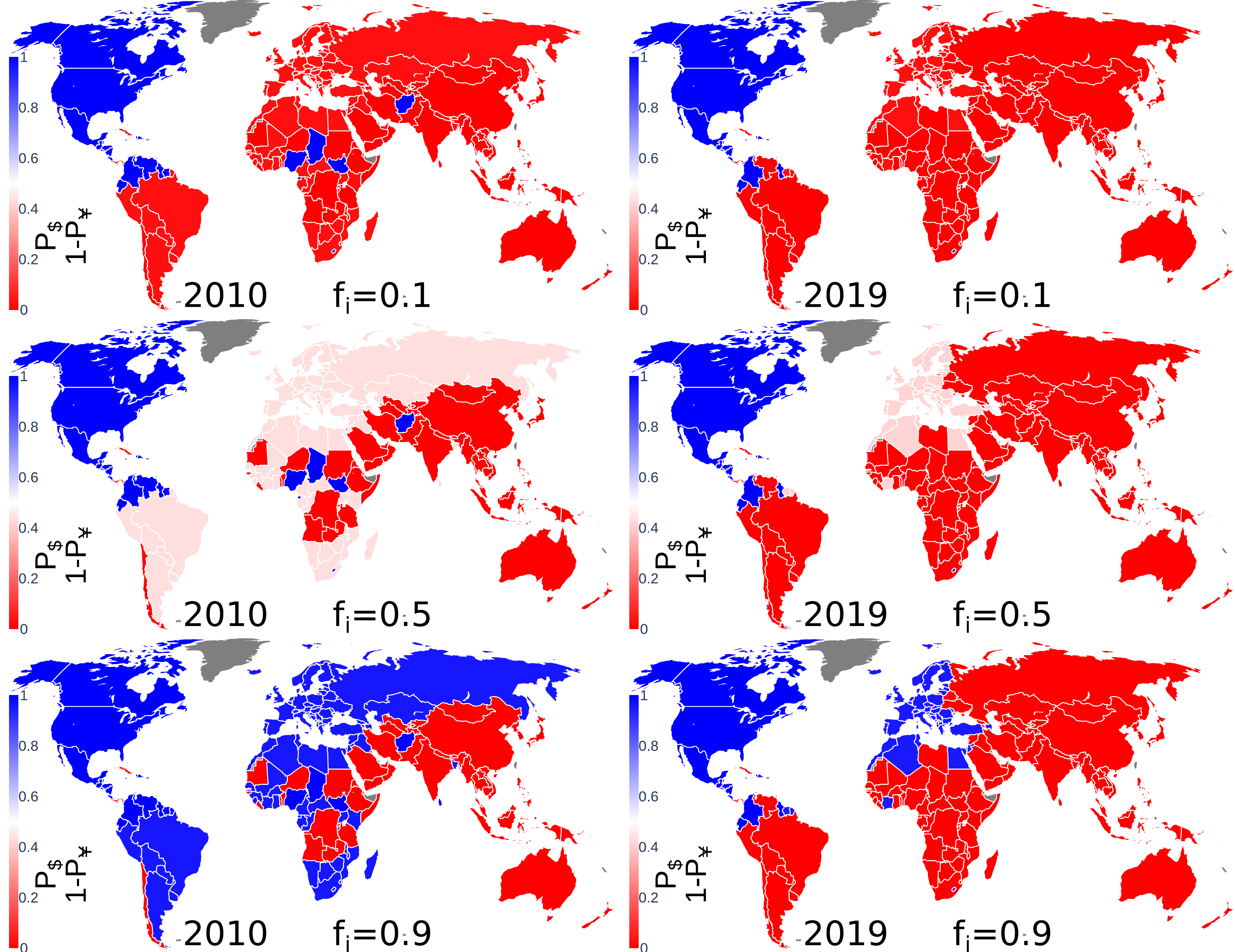}
	\end{center}
	\vglue -0.3cm
	\caption{\label{fig3}World distribution of the trade currency preference probability for the years 2010 and 2019. Each panel corresponds to a given year and a given fraction $f_i$ of countries preferring to initially trade in USD. Left (right) column panels correspond to the year 2010 (2019). The top, middle, and bottom rows correspond to $f_i=0.1$, $0.5$, and $0.9$, respectively. Each country is characterized by a probability $P_\$$ to obtain a USD trade preference at the end of the Monte Carlo procedure. The CNY trade preference probability of a country is then $P_\yen=1-P_\$$. The color ranges from red for $P_\$=0$ and $P_\yen=1$ to blue for $P_\$=1$ and $P_\yen=0$. The average TCP probability $P_\$$ has been computed from $N_r=10^4$ random initial TCP distributions.
	}
\end{figure}

\begin{figure}[t]
	\begin{center}
		\includegraphics[width=\columnwidth]{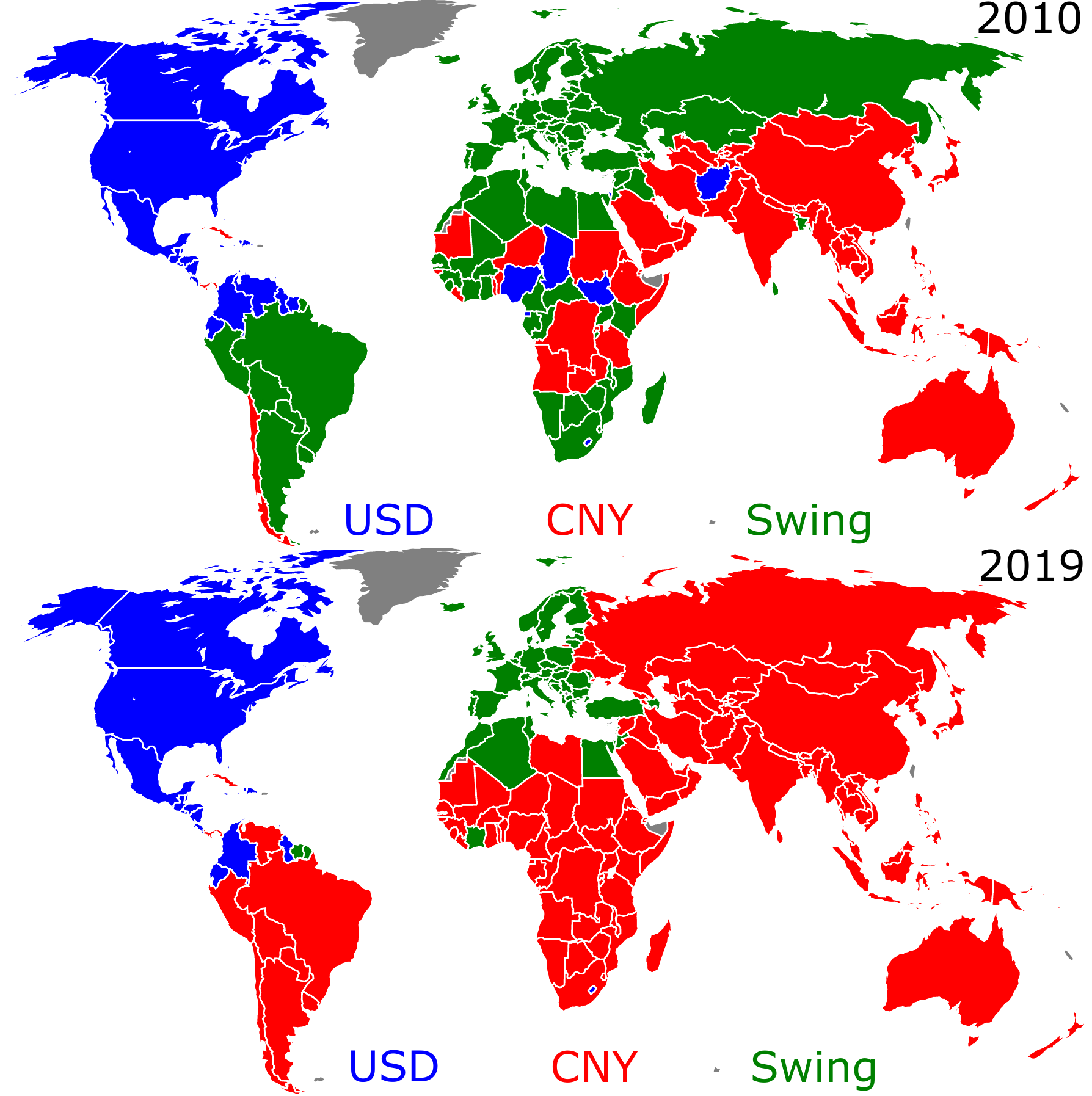}
	\end{center}
	\vglue -0.3cm
	\caption{\label{fig4}
		World distribution of countries belonging to the USD group
		(blue, hard preference to trade in USD), the CNY group (red, hard preference to trade in CNY)
		and the swing group (green, the TCP can change between USD and CNY depending on the initial conditions). The world maps are shown for the years 2010 (left panel) and 2019 (right panel).
	}
\end{figure}

Let us define the TCP probability $P_\$(c)$ to obtain for the country $c$ a USD preference at the end of the Monte Carlo procedure. The probability to obtain for the country $c$ a CNY preference is then $P_\yen(c)=1-P_\$(c)$. The probability $P_\$$ is obtained from the application of the Monte Carlo procedure to the $N_r=10^4$ initial TCP distributions. Fig.~\ref{fig3} shows the TCP probability world distribution. The Fig.~\ref{fig3} left panels illustrate the above described bipolar USD-CNY trade currency preference which exists in 2010: for high (low) $f_i$ most of the countries finally prefer USD (CNY). In 2019, Fig.~\ref{fig3} right panels, the CNY dominance is clearly observed. Indeed, even for high $f_i$, most of the countries finally prefer CNY over USD.

The reason of the bistability of the final outcomes $f_f$ (see $f_1$ and $f_2$ in Fig.~\ref{fig2})
can be understood from the analysis of the distribution of the USD or CNY trade preference over the
countries. In fact, there are two groups of countries
which keep, for any initial fraction $f_i$, a hard preference to trade in USD
(the USD group) and in CNY (the CNY group). Otherwise stated, a country of the CNY (USD) group, independently of its initial TCP and of the initial TCPs of the other countries, will always ends up in the CNY (USD) group.
A third group of swing states (the swing group) may change their TCP. Amazingly, 
depending on the initial configuration of countries which prefer to trade in USD or in CNY, the countries belonging to this swing group collectively adopt at the equilibrium either the USD or the CNY as trade currency. This swing states are collectively responsible for the final outcome: if they adopt USD (CNY) at the equilibrium, the final fraction $f_f$ will be the highest (lowest) of the two possible fractions, i.e., $f_2$ ($f_1$).
These three groups are shown on the world maps displayed in Fig.~\ref{fig4} for the years 2010 and 2019
(the world maps for the other years of the past decade are shown in Fig.~\ref{figS3} of the Supplementary Information).
We clearly see that there is a drastic change from 2010 to 2019:
a large number of countries passed from the swing group to the CNY group which has considerably increased in size during the last decade. Indeed,  the former Soviet Union countries,
almost all South America and Africa belong in 2019 to the CNY group.
By contrast, from 2010 to 2019, the size of the USD group reduced only slightly loosing few countries: Venezuela,  Nigeria, Chad, South Sudan, Equatorial Guinea, Afghanistan and Federated States of Micronesia switched to the CNY group and, Suriname and Israel to the swing group.
In 2019, the swing group is mainly composed by
EU countries, the UK and some Mediterranean countries (Turkey, Egypt, Morocco, Algeria, Tunisia and Israel).
The lists of the countries belonging to the USD, CNY and swing groups in 2010 and in 2019 are given in Tables~\ref{tab:tabS1} to \ref{tab:tabS6} of the Supplementary Information.

\begin{figure}[t]
	\begin{center}
		\includegraphics[width=\columnwidth]{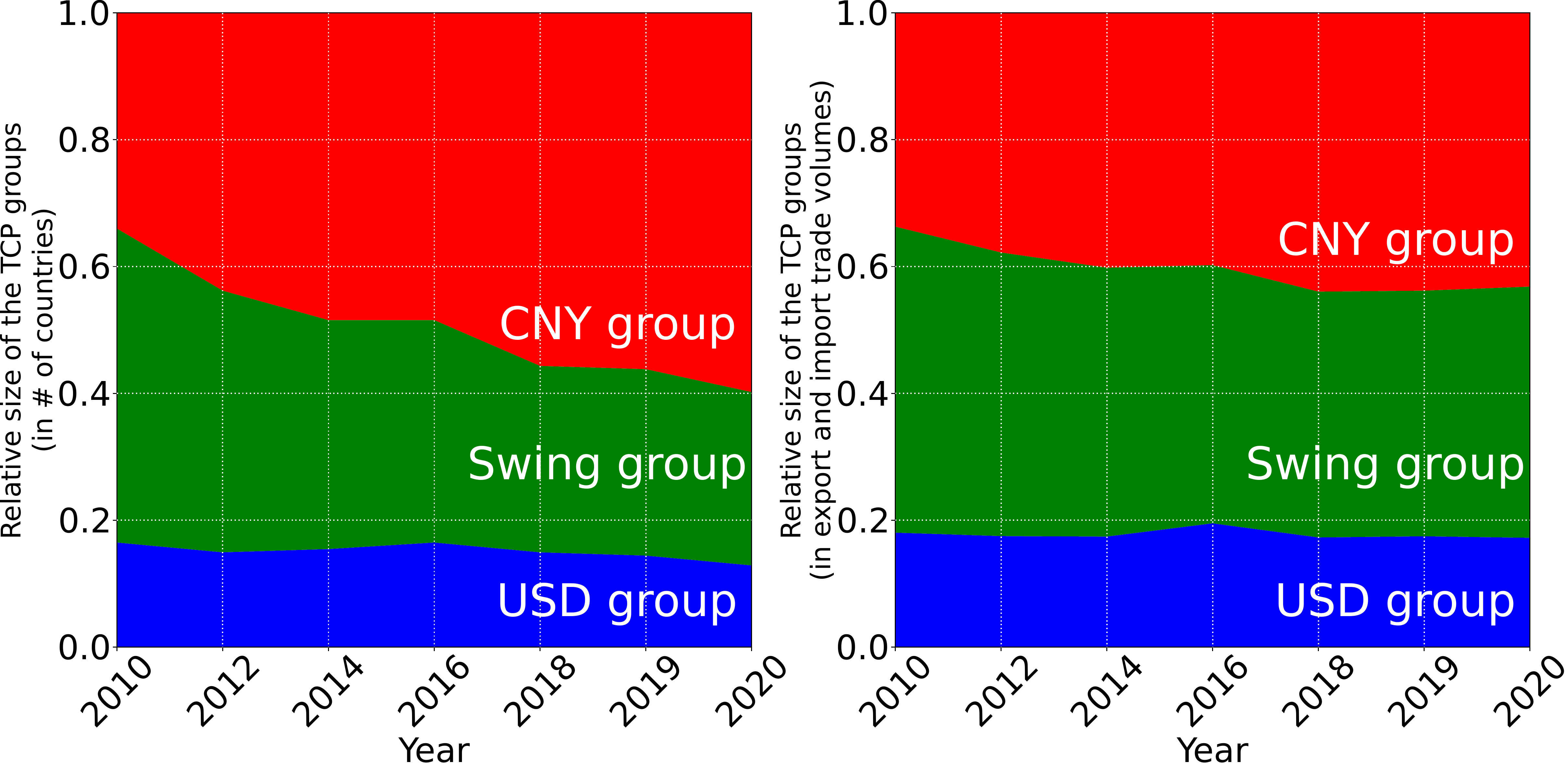}
	\end{center}
	\vglue -0.3cm
	\caption{\label{fig5}Time evolution of the sizes of the trade currency preference groups. The 3 bands correspond to the USD group (blue), the CNY group (red), and the swing group (green). The width of a band corresponds to the size of the corresponding group expressed as:
		(left panel) the ratio between the countries belonging to the group over and the total number of countries,
		(right panel) the ratio between the total trade volume exchanged by the countries of the group and the total volume exchanged by all the world countries. The trade volumes are expressed in USD of the concerned year.}
\end{figure}

The time evolution over the last decade of the size of the three TCP groups is shown in Fig.~\ref{fig5}. The left panel of Fig.~\ref{fig5} displays the fraction of countries belonging to each group and the right panel displays the fraction of the total volume of import and export exchanged by each group. From Fig.~\ref{fig5} left panel, we observe that the CNY group (red band) steadily grows along the decade from a fraction of 34\% of the countries in 2010 to 60\% in 2020. This growth of the CNY group is mainly compensated by the depletion of the swing group (green band) whose size drops from 50\% of the countries in 2010 to 27\% in 2020. Meanwhile, the size of the USD group (blue band) slightly decreased from 16\% of the countries in 2010 to 13\% in 2020.
The trends are the same but less pronounced for the fraction of the trade volume exchanged by the different groups (see Fig.~\ref{fig5} right panel). The fraction of the trade volume exchanged by the CNY (swing) group increases (decreases) from 34\% (49\%) in 2010 to 43\% (40\%) in 2020. Meanwhile, the fraction of the trade volume exchanged by the USD group stayed quite constant during the decade (17\% for both 2010 and 2020). Consequently,  the number of countries switching during the last decade from the swing group to the CNY group represents 23\% of the world countries but represents only 9\% of the world trade volume. The CNY club increased but with somewhat less important new entrants in terms of trade volume exchanged.

Let us note that we obtain practically the same results if, instead of keeping China always trading in CNY and the USA always trading in USD, we keep China and the other BRICS (Brazil, Russia, India, South Africa) always trading in CNY and the USA and other Anglo-Saxon countries (Canada, UK, Australia, New Zealand) always trading in USD; see e.g., Figs.~\ref{figS4} and \ref{figS5} in the Supplementary Information which are quite similar to Figs.~\ref{fig2} and \ref{fig4}. This result asserts the dominance of China and USA in the world trade network.
We have also considered to replace the import trade probabilities $P_c$ and the export trade probabilities $P^*_c$ in
(\ref{eq1}) by the PageRank and the CheiRank probabilities
obtained from the Google matrix of the WTN. These probabilities allowing to measure the capability of a country to import or export products throughout the WTN were used to analyze the international trade \cite{wtn1,wtn3,wtncrisis}. Such a replacement of the probabilities, i.e., of the centrality measures of the WTN,
leads again to practically the same results (see e.g., Figs.~\ref{figS6} and \ref{figS7} in the Supplementary Information which are similar to Figs.~\ref{fig2} and \ref{fig4}).

\section{Conclusion and discussion}
The question addressed in the current work is the following one: assuming that only the WTN structure matters, what would be the trade currency preference for each country?
As a first step we used a model with two currencies associated to the nowadays leading economies, i.e., USA and China. A next step toward a more refined model would be to consider additional currencies such as the EUR for the eurozone. The world economy being strongly polarized, it is illusory to consider more than three dominant trade currencies. The world partition obtained by a naive application of the Louvain modularity method on the bare WTN supports this assertions (see Fig.~\ref{fig1}). Whether or not the swing group, which in 2019 is reduced to the eurozone and other EUR dependent economies (see Fig.~\ref{fig4}), crystallizes into an stable EUR group is an open question. The possible stability of a three trade currency model is another question. We leave these questions for a subsequent work. The results presented here for the CNY-USD trade currency model should capture main features of more refined models.

As a conclusion, our analysis, performed by superimposing an Ising spin network on the WTN, clearly shows that the structure of the international trade would favor nowadays the main part of the world to trade in CNY while in 2010 it would have favored the trade in USD.

We observe two final equilibrium states. In 2010, one of them characterizes  an  USD preference and the other one a CNY preference, whilst in 2019, both of the two final states characterize a CNY preference. Nowadays, according to the WTN structure, for any initial distributions of countries preferring to trade in USD and in CNY, the final state would always favor a world which preferentially trade in CNY. The bistability of the final state is due to a group of swing states which, depending on the initial distribution of the trade currency preferences over the countries, adopt all together a preference for either USD or CNY. Of course, our analysis is based on the mathematical treatment
of the trade flows between the world countries and does not take
into account any geopolitical relations between the countries.
But, it is often claimed that economics determines politics and thus
we argue that the obtained results demonstrate drastic changes in the international trade structure which favors now yuan over dollar.

These results obtained from the intertwined structure of the international trade flows echo the current questioning about a hypothetical replacement of USD by CNY as the global currency \cite{FP,BS,bloomberg,balance,liu22,carnegie} and the current trends consisting to label contracts in CNY for Saudi Arabia to China or Russia to China crude oil and petrol imports \cite{liu22,nikkei,wallstrj,oilprice,bloomberg2}. Although the road to internationalization of the CNY is still long \cite{liu19}  and although some serious criteria, such as  the transparency of China’s financial markets and
the perceived-from-abroad-stability of the Chinese monetary policies, are still lacking to turn CNY into a global currency, our results show nonetheless that the international trade network is ready to harbor the USD vs CNY competition and the possibly USD to CNY transition. Hence, as the nowadays global trade flows structure favors CNY, interested economic stakeholders could be encouraged to bypass USD in order to establish contract in CNY for a large variety of economic sectors going \textit{de facto} beyond the current niche use of CNY in the crude oil and gas market.

%%%%%%%%%%%%%%%%%%%%%%%%%%%%%%%%%%%%%%%%%%
\vspace{6pt} 

%%%%%%%%%%%%%%%%%%%%%%%%%%%%%%%%%%%%%%%%%%
%% optional
%\supplementary{The following supporting information can be downloaded at:  \linksupplementary{s1}, Figure S1: title; Table S1: title; Video S1: title.}


%%%%%%%%%%%%%%%%%%%%%%%%%%%%%%%%%%%%%%%%%%
\begin{acknowledgments}
We thank L. Ermann for his help to collect data from the UN Comtrade database. We thank the UN Statistics Division to grant us a friendly access to the UN Comtrade database. This research has been partially supported by the grant
	NANOX N$^\circ$ ANR-17-EURE-0009 (project MTDINA) in the frame 
	of the Programme des Investissements d'Avenir, France. 
	This research has also been supported by the
	Programme Investissements d’Avenir ANR-15-IDEX-0003.
\end{acknowledgments}
\section*{abbreviation}
\begin{tabular}{@{}ll}
CNY&Chinese yuan\\
TCP&Trade currency preference\\
UN& United Nations\\
USD& United States dollar\\
WTN& World trade network
\end{tabular}
% Bibliography
\clearpage
\bibliography{dywtnARXIV}

\begin{thebibliography}{37}%
\makeatletter
\providecommand \@ifxundefined [1]{%
 \@ifx{#1\undefined}
}%
\providecommand \@ifnum [1]{%
 \ifnum #1\expandafter \@firstoftwo
 \else \expandafter \@secondoftwo
 \fi
}%
\providecommand \@ifx [1]{%
 \ifx #1\expandafter \@firstoftwo
 \else \expandafter \@secondoftwo
 \fi
}%
\providecommand \natexlab [1]{#1}%
\providecommand \enquote  [1]{``#1''}%
\providecommand \bibnamefont  [1]{#1}%
\providecommand \bibfnamefont [1]{#1}%
\providecommand \citenamefont [1]{#1}%
\providecommand \href@noop [0]{\@secondoftwo}%
\providecommand \href [0]{\begingroup \@sanitize@url \@href}%
\providecommand \@href[1]{\@@startlink{#1}\@@href}%
\providecommand \@@href[1]{\endgroup#1\@@endlink}%
\providecommand \@sanitize@url [0]{\catcode `\\12\catcode `\$12\catcode
  `\&12\catcode `\#12\catcode `\^12\catcode `\_12\catcode `\%12\relax}%
\providecommand \@@startlink[1]{}%
\providecommand \@@endlink[0]{}%
\providecommand \url  [0]{\begingroup\@sanitize@url \@url }%
\providecommand \@url [1]{\endgroup\@href {#1}{\urlprefix }}%
\providecommand \urlprefix  [0]{URL }%
\providecommand \Eprint [0]{\href }%
\providecommand \doibase [0]{https://doi.org/}%
\providecommand \selectlanguage [0]{\@gobble}%
\providecommand \bibinfo  [0]{\@secondoftwo}%
\providecommand \bibfield  [0]{\@secondoftwo}%
\providecommand \translation [1]{[#1]}%
\providecommand \BibitemOpen [0]{}%
\providecommand \bibitemStop [0]{}%
\providecommand \bibitemNoStop [0]{.\EOS\space}%
\providecommand \EOS [0]{\spacefactor3000\relax}%
\providecommand \BibitemShut  [1]{\csname bibitem#1\endcsname}%
\let\auto@bib@innerbib\@empty
%</preamble>
\bibitem [{\citenamefont {{Wikipedia contributors}}(2022)}]{wiki1}%
  \BibitemOpen
  \bibfield  {author} {\bibinfo {author} {\bibnamefont {{Wikipedia
  contributors}}},\ }\href
  {https://en.wikipedia.org/w/index.php?title=Bretton_Woods_system&oldid=1114289262}
  {\bibinfo {title} {{Bretton Woods system --- {Wikipedia}{,} The Free
  Encyclopedia}}} (\bibinfo {year} {2022})\BibitemShut {NoStop}%
\bibitem [{\citenamefont {{United Nations Statistics Division}}()}]{comtrade}%
  \BibitemOpen
  \bibfield  {author} {\bibinfo {author} {\bibnamefont {{United Nations
  Statistics Division}}},\ }\href {http://comtrade.un.org/db/} {\bibinfo
  {title} {United nations commodity trade statistics database}}\BibitemShut
  {NoStop}%
\bibitem [{\citenamefont {Raisinghani}()}]{FP}%
  \BibitemOpen
  \bibfield  {author} {\bibinfo {author} {\bibfnamefont {V.}~\bibnamefont
  {Raisinghani}},\ }\href
  {https://financialpost.com/moneywise/could-chinas-yuan-replace-the-u-s-dollar-as-the-worlds-dominant-currency}
  {\bibinfo {title} {{Could China's Yuan replace the U.S. dollar as the world's
  dominant currency?}}},\ \bibinfo {note} {{Financial Post}}\BibitemShut
  {NoStop}%
\bibitem [{\citenamefont {Ahmed}()}]{BS}%
  \BibitemOpen
  \bibfield  {author} {\bibinfo {author} {\bibfnamefont {S.~R.}\ \bibnamefont
  {Ahmed}},\ }\href
  {https://www.tbsnews.net/features/panorama/can-yuan-replace-mighty-dollar-reserve-currency-497706}
  {\bibinfo {title} {{Can yuan replace the ‘mighty’ dollar?}}},\ \bibinfo
  {note} {{The Business Standard}}\BibitemShut {NoStop}%
\bibitem [{\citenamefont {{Global Times writers}}()}]{GT}%
  \BibitemOpen
  \bibfield  {author} {\bibinfo {author} {\bibnamefont {{Global Times
  writers}}},\ }\href {https://www.globaltimes.cn/page/202210/1276588.shtml}
  {\bibinfo {title} {{Chinese yuan becomes most traded foreign currency on the
  Moscow Exchange, surpasses the US dollar: report}}},\ \bibinfo {note}
  {{Global Times}}\BibitemShut {NoStop}%
\bibitem [{\citenamefont {Amadeo}()}]{balance}%
  \BibitemOpen
  \bibfield  {author} {\bibinfo {author} {\bibfnamefont {K.}~\bibnamefont
  {Amadeo}},\ }\href
  {https://www.thebalancemoney.com/yuan-reserve-currency-to-global-currency-3970465}
  {\bibinfo {title} {{How the Yuan Could Become a Global Currency}}},\ \bibinfo
  {note} {{The balance}}\BibitemShut {NoStop}%
\bibitem [{\citenamefont {Curran}()}]{bloomberg}%
  \BibitemOpen
  \bibfield  {author} {\bibinfo {author} {\bibfnamefont {E.}~\bibnamefont
  {Curran}},\ }\href
  {https://www.bloomberg.com/news/articles/2022-03-25/the-dollar-s-dominance-is-being-stealthily-eroded-imf-paper}
  {\bibinfo {title} {{The U.S. Dollar’s Dominance Is Being Stealthily
  Eroded}}},\ \bibinfo {note} {bloomberg}\BibitemShut {NoStop}%
\bibitem [{\citenamefont {Liu}\ and\ \citenamefont {Papa}(2022)}]{liu22}%
  \BibitemOpen
  \bibfield  {author} {\bibinfo {author} {\bibfnamefont {Z.~Z.}\ \bibnamefont
  {Liu}}\ and\ \bibinfo {author} {\bibfnamefont {M.}~\bibnamefont {Papa}},\
  }\href {https://doi.org/10.1017/9781009029544} {\emph {\bibinfo {title} {Can
  BRICS De-dollarize the Global Financial System?}}},\ Elements in the
  Economics of Emerging Markets\ (\bibinfo  {publisher} {Cambridge University
  Press},\ \bibinfo {year} {2022})\BibitemShut {NoStop}%
\bibitem [{\citenamefont {{Nikkei staff writers}}()}]{nikkei}%
  \BibitemOpen
  \bibfield  {author} {\bibinfo {author} {\bibnamefont {{Nikkei staff
  writers}}},\ }\href
  {https://asia.nikkei.com/Business/Markets/Currencies/Russian-companies-shift-to-yuan-as-flight-from-dollar-accelerates}
  {\bibinfo {title} {{Russian companies shift to yuan as flight from dollar
  accelerates}}},\ \bibinfo {note} {{Nikkei Asia}}\BibitemShut {NoStop}%
\bibitem [{\citenamefont {Said}\ and\ \citenamefont {Kalin}()}]{wallstrj}%
  \BibitemOpen
  \bibfield  {author} {\bibinfo {author} {\bibfnamefont {S.}~\bibnamefont
  {Said}}\ and\ \bibinfo {author} {\bibfnamefont {S.}~\bibnamefont {Kalin}},\
  }\href
  {https://www.wsj.com/articles/saudi-arabia-considers-accepting-yuan-instead-of-dollars-for-chinese-oil-sales-11647351541}
  {\bibinfo {title} {{Saudi Arabia Considers Accepting Yuan Instead of Dollars
  for Chinese Oil Sales}}},\ \bibinfo {note} {{The Wall Street
  Journal}}\BibitemShut {NoStop}%
\bibitem [{\citenamefont {{The World Bank}}()}]{chinaecon}%
  \BibitemOpen
  \bibfield  {author} {\bibinfo {author} {\bibnamefont {{The World Bank}}},\
  }\href {https://www.worldbank.org/en/country/china/overview} {\bibinfo
  {title} {{China Overview: Development news, research, data}}}\BibitemShut
  {NoStop}%
\bibitem [{\citenamefont {{World Trade Organization}}()}]{WTOSR22}%
  \BibitemOpen
  \bibfield  {author} {\bibinfo {author} {\bibnamefont {{World Trade
  Organization}}},\ }\href
  {https://www.wto.org/english/res_e/publications_e/wtsr_2022_e.htm} {\bibinfo
  {title} {{World Trade Statistical Review 2022}}}\BibitemShut {NoStop}%
\bibitem [{\citenamefont {Krugman}\ \emph {et~al.}(2011)\citenamefont
  {Krugman}, \citenamefont {Obstfeld},\ and\ \citenamefont
  {Melitz}}]{krugman11}%
  \BibitemOpen
  \bibfield  {author} {\bibinfo {author} {\bibfnamefont {P.}~\bibnamefont
  {Krugman}}, \bibinfo {author} {\bibfnamefont {M.}~\bibnamefont {Obstfeld}},\
  and\ \bibinfo {author} {\bibfnamefont {M.}~\bibnamefont {Melitz}},\
  }\href@noop {} {\emph {\bibinfo {title} {{International Economics: Theory \&
  Policy}}}}\ (\bibinfo  {publisher} {Prentice Hall, New Jersey},\ \bibinfo
  {year} {2011})\BibitemShut {NoStop}%
\bibitem [{\citenamefont {Galam}(2008)}]{galam08}%
  \BibitemOpen
  \bibfield  {author} {\bibinfo {author} {\bibfnamefont {S.}~\bibnamefont
  {Galam}},\ }\bibfield  {title} {\bibinfo {title} {Sociophysics: a review of
  galam models},\ }\href {https://doi.org/10.1142/S0129183108012297} {\bibfield
   {journal} {\bibinfo  {journal} {International Journal of Modern Physics C}\
  }\textbf {\bibinfo {volume} {19}},\ \bibinfo {pages} {409} (\bibinfo {year}
  {2008})},\ \Eprint
  {https://arxiv.org/abs/https://doi.org/10.1142/S0129183108012297}
  {https://doi.org/10.1142/S0129183108012297} \BibitemShut {NoStop}%
\bibitem [{\citenamefont {Castellano}\ \emph {et~al.}(2009)\citenamefont
  {Castellano}, \citenamefont {Fortunato},\ and\ \citenamefont
  {Loreto}}]{castellano09}%
  \BibitemOpen
  \bibfield  {author} {\bibinfo {author} {\bibfnamefont {C.}~\bibnamefont
  {Castellano}}, \bibinfo {author} {\bibfnamefont {S.}~\bibnamefont
  {Fortunato}},\ and\ \bibinfo {author} {\bibfnamefont {V.}~\bibnamefont
  {Loreto}},\ }\bibfield  {title} {\bibinfo {title} {Statistical physics of
  social dynamics},\ }\href {https://doi.org/10.1103/RevModPhys.81.591}
  {\bibfield  {journal} {\bibinfo  {journal} {Rev. Mod. Phys.}\ }\textbf
  {\bibinfo {volume} {81}},\ \bibinfo {pages} {591} (\bibinfo {year}
  {2009})}\BibitemShut {NoStop}%
\bibitem [{\citenamefont {Kandiah}\ and\ \citenamefont
  {Shepelyansky}(2012)}]{opinion}%
  \BibitemOpen
  \bibfield  {author} {\bibinfo {author} {\bibfnamefont {V.}~\bibnamefont
  {Kandiah}}\ and\ \bibinfo {author} {\bibfnamefont {D.}~\bibnamefont
  {Shepelyansky}},\ }\bibfield  {title} {\bibinfo {title} {{PageRank model of
  opinion formation on social networks}},\ }\href
  {https://doi.org/10.1016/j.physa.2012.06.047} {\bibfield  {journal} {\bibinfo
   {journal} {{Physica A}}\ }\textbf {\bibinfo {volume} {391}},\ \bibinfo
  {pages} {5779} (\bibinfo {year} {2012})}\BibitemShut {NoStop}%
\bibitem [{\citenamefont {Schmittmann}\ and\ \citenamefont
  {Mukhopadhyay}(2010)}]{schmittmann10}%
  \BibitemOpen
  \bibfield  {author} {\bibinfo {author} {\bibfnamefont {B.}~\bibnamefont
  {Schmittmann}}\ and\ \bibinfo {author} {\bibfnamefont {A.}~\bibnamefont
  {Mukhopadhyay}},\ }\bibfield  {title} {\bibinfo {title} {Opinion formation on
  adaptive networks with intensive average degree},\ }\href
  {https://doi.org/10.1103/PhysRevE.82.066104} {\bibfield  {journal} {\bibinfo
  {journal} {Phys. Rev. E}\ }\textbf {\bibinfo {volume} {82}},\ \bibinfo
  {pages} {066104} (\bibinfo {year} {2010})}\BibitemShut {NoStop}%
\bibitem [{\citenamefont {Dorogovtsev}(2010)}]{dorogovtsev}%
  \BibitemOpen
  \bibfield  {author} {\bibinfo {author} {\bibfnamefont {S.}~\bibnamefont
  {Dorogovtsev}},\ }\href
  {https://doi.org/10.1093/acprof:oso/9780199548927.001.0001} {\emph {\bibinfo
  {title} {{Lectures in Complex Networks}}}}\ (\bibinfo  {publisher} {Oxford
  University Press},\ \bibinfo {year} {2010})\BibitemShut {NoStop}%
\bibitem [{\citenamefont {Langville}\ and\ \citenamefont
  {Meyer}(2006)}]{meyer}%
  \BibitemOpen
  \bibfield  {author} {\bibinfo {author} {\bibfnamefont {A.}~\bibnamefont
  {Langville}}\ and\ \bibinfo {author} {\bibfnamefont {C.}~\bibnamefont
  {Meyer}},\ }\href@noop {} {\emph {\bibinfo {title} {{Google's PageRank and
  beyond: the science of search engine rankings}}}}\ (\bibinfo  {publisher}
  {Princeton University Press},\ \bibinfo {year} {2006})\BibitemShut {NoStop}%
\bibitem [{\citenamefont {Ermann}\ \emph {et~al.}(2015)\citenamefont {Ermann},
  \citenamefont {Frahm},\ and\ \citenamefont {Shepelyansky}}]{rmp2015}%
  \BibitemOpen
  \bibfield  {author} {\bibinfo {author} {\bibfnamefont {L.}~\bibnamefont
  {Ermann}}, \bibinfo {author} {\bibfnamefont {K.}~\bibnamefont {Frahm}},\ and\
  \bibinfo {author} {\bibfnamefont {D.}~\bibnamefont {Shepelyansky}},\
  }\bibfield  {title} {\bibinfo {title} {{Google matrix analysis of directed
  networks}},\ }\href {https://doi.org/10.1103/RevModPhys.87.1261} {\bibfield
  {journal} {\bibinfo  {journal} {{Rev. Mod. Phys.}}\ }\textbf {\bibinfo
  {volume} {87}},\ \bibinfo {pages} {1261} (\bibinfo {year}
  {2015})}\BibitemShut {NoStop}%
\bibitem [{\citenamefont {Serrano}\ \emph {et~al.}(2007)\citenamefont
  {Serrano}, \citenamefont {Boguñá},\ and\ \citenamefont
  {Vespignani}}]{serrano07}%
  \BibitemOpen
  \bibfield  {author} {\bibinfo {author} {\bibfnamefont {M.~A.}\ \bibnamefont
  {Serrano}}, \bibinfo {author} {\bibfnamefont {M.}~\bibnamefont {Boguñá}},\
  and\ \bibinfo {author} {\bibfnamefont {A.}~\bibnamefont {Vespignani}},\
  }\bibfield  {title} {\bibinfo {title} {Patterns of dominant flows in the
  world trade web},\ }\href {https://doi.org/10.1007/s11403-007-0026-y}
  {\bibfield  {journal} {\bibinfo  {journal} {{Journal of Economic Interaction
  and Coordination}}\ }\textbf {\bibinfo {volume} {2}},\ \bibinfo {pages} {111}
  (\bibinfo {year} {2007})}\BibitemShut {NoStop}%
\bibitem [{\citenamefont {Fagiolo}\ \emph {et~al.}(2009)\citenamefont
  {Fagiolo}, \citenamefont {Reyes},\ and\ \citenamefont {Schiavo}}]{fagiolo09}%
  \BibitemOpen
  \bibfield  {author} {\bibinfo {author} {\bibfnamefont {G.}~\bibnamefont
  {Fagiolo}}, \bibinfo {author} {\bibfnamefont {J.}~\bibnamefont {Reyes}},\
  and\ \bibinfo {author} {\bibfnamefont {S.}~\bibnamefont {Schiavo}},\
  }\bibfield  {title} {\bibinfo {title} {{World-trade web: Topological
  properties, dynamics, and evolution}},\ }\href
  {https://doi.org/10.1103/PhysRevE.79.036115} {\bibfield  {journal} {\bibinfo
  {journal} {Phys. Rev. E}\ }\textbf {\bibinfo {volume} {79}},\ \bibinfo
  {pages} {036115} (\bibinfo {year} {2009})}\BibitemShut {NoStop}%
\bibitem [{\citenamefont {He}\ and\ \citenamefont {Deem}(2010)}]{he10}%
  \BibitemOpen
  \bibfield  {author} {\bibinfo {author} {\bibfnamefont {J.}~\bibnamefont
  {He}}\ and\ \bibinfo {author} {\bibfnamefont {M.~W.}\ \bibnamefont {Deem}},\
  }\bibfield  {title} {\bibinfo {title} {{Structure and Response in the World
  Trade Network}},\ }\href {https://doi.org/10.1103/PhysRevLett.105.198701}
  {\bibfield  {journal} {\bibinfo  {journal} {Phys. Rev. Lett.}\ }\textbf
  {\bibinfo {volume} {105}},\ \bibinfo {pages} {198701} (\bibinfo {year}
  {2010})}\BibitemShut {NoStop}%
\bibitem [{\citenamefont {Fagiolo}\ \emph {et~al.}(2010)\citenamefont
  {Fagiolo}, \citenamefont {Reyes},\ and\ \citenamefont {Schiavo}}]{fagiolo10}%
  \BibitemOpen
  \bibfield  {author} {\bibinfo {author} {\bibfnamefont {G.}~\bibnamefont
  {Fagiolo}}, \bibinfo {author} {\bibfnamefont {J.}~\bibnamefont {Reyes}},\
  and\ \bibinfo {author} {\bibfnamefont {S.}~\bibnamefont {Schiavo}},\
  }\bibfield  {title} {\bibinfo {title} {The evolution of the world trade web:
  a weighted-network analysis},\ }\href
  {https://doi.org/10.1007/s00191-009-0160-x} {\bibfield  {journal} {\bibinfo
  {journal} {Journal of Evolutionary Economics}\ }\textbf {\bibinfo {volume}
  {20}},\ \bibinfo {pages} {479} (\bibinfo {year} {2010})}\BibitemShut
  {NoStop}%
\bibitem [{\citenamefont {Barigozzi}\ \emph {et~al.}(2010)\citenamefont
  {Barigozzi}, \citenamefont {Fagiolo},\ and\ \citenamefont
  {Garlaschelli}}]{barigozzi10}%
  \BibitemOpen
  \bibfield  {author} {\bibinfo {author} {\bibfnamefont {M.}~\bibnamefont
  {Barigozzi}}, \bibinfo {author} {\bibfnamefont {G.}~\bibnamefont {Fagiolo}},\
  and\ \bibinfo {author} {\bibfnamefont {D.}~\bibnamefont {Garlaschelli}},\
  }\bibfield  {title} {\bibinfo {title} {Multinetwork of international trade: A
  commodity-specific analysis},\ }\href
  {https://doi.org/10.1103/PhysRevE.81.046104} {\bibfield  {journal} {\bibinfo
  {journal} {Phys. Rev. E}\ }\textbf {\bibinfo {volume} {81}},\ \bibinfo
  {pages} {046104} (\bibinfo {year} {2010})}\BibitemShut {NoStop}%
\bibitem [{\citenamefont {Chakraborty}\ \emph {et~al.}(2018)\citenamefont
  {Chakraborty}, \citenamefont {Kichikawa}, \citenamefont {Iino}, \citenamefont
  {Iyetomi}, \citenamefont {Inoue}, \citenamefont {Fujiwara},\ and\
  \citenamefont {Aoyama}}]{chakraborty18}%
  \BibitemOpen
  \bibfield  {author} {\bibinfo {author} {\bibfnamefont {A.}~\bibnamefont
  {Chakraborty}}, \bibinfo {author} {\bibfnamefont {Y.}~\bibnamefont
  {Kichikawa}}, \bibinfo {author} {\bibfnamefont {T.}~\bibnamefont {Iino}},
  \bibinfo {author} {\bibfnamefont {H.}~\bibnamefont {Iyetomi}}, \bibinfo
  {author} {\bibfnamefont {H.}~\bibnamefont {Inoue}}, \bibinfo {author}
  {\bibfnamefont {Y.}~\bibnamefont {Fujiwara}},\ and\ \bibinfo {author}
  {\bibfnamefont {H.}~\bibnamefont {Aoyama}},\ }\bibfield  {title} {\bibinfo
  {title} {Hierarchical communities in the walnut structure of the japanese
  production network},\ }\href {https://doi.org/10.1371/journal.pone.0202739}
  {\bibfield  {journal} {\bibinfo  {journal} {PLOS ONE}\ }\textbf {\bibinfo
  {volume} {13}},\ \bibinfo {pages} {1} (\bibinfo {year} {2018})}\BibitemShut
  {NoStop}%
\bibitem [{\citenamefont {De~Benedictis}\ and\ \citenamefont
  {Tajoli}(2011)}]{debenedictis11}%
  \BibitemOpen
  \bibfield  {author} {\bibinfo {author} {\bibfnamefont {L.}~\bibnamefont
  {De~Benedictis}}\ and\ \bibinfo {author} {\bibfnamefont {L.}~\bibnamefont
  {Tajoli}},\ }\bibfield  {title} {\bibinfo {title} {The world trade network},\
  }\href {https://doi.org/https://doi.org/10.1111/j.1467-9701.2011.01360.x}
  {\bibfield  {journal} {\bibinfo  {journal} {The World Economy}\ }\textbf
  {\bibinfo {volume} {34}},\ \bibinfo {pages} {1417} (\bibinfo {year}
  {2011})},\ \Eprint
  {https://arxiv.org/abs/https://onlinelibrary.wiley.com/doi/pdf/10.1111/j.1467-9701.2011.01360.x}
  {https://onlinelibrary.wiley.com/doi/pdf/10.1111/j.1467-9701.2011.01360.x}
  \BibitemShut {NoStop}%
\bibitem [{\citenamefont {Ermann}\ and\ \citenamefont
  {Shepelyansky}(2011)}]{wtn1}%
  \BibitemOpen
  \bibfield  {author} {\bibinfo {author} {\bibfnamefont {L.}~\bibnamefont
  {Ermann}}\ and\ \bibinfo {author} {\bibfnamefont {D.}~\bibnamefont
  {Shepelyansky}},\ }\bibfield  {title} {\bibinfo {title} {{Google matrix of
  the world trade network}},\ }\href
  {https://doi.org/10.12693/APhysPolA.120.A-158} {\bibfield  {journal}
  {\bibinfo  {journal} {{Acta Physica Polonica A}}\ }\textbf {\bibinfo {volume}
  {120}},\ \bibinfo {pages} {A158} (\bibinfo {year} {2011})}\BibitemShut
  {NoStop}%
\bibitem [{\citenamefont {Coquidé}\ \emph {et~al.}(2019)\citenamefont
  {Coquidé}, \citenamefont {Ermann}, \citenamefont {Lages},\ and\
  \citenamefont {Shepelyansky}}]{wtn3}%
  \BibitemOpen
  \bibfield  {author} {\bibinfo {author} {\bibfnamefont {C.}~\bibnamefont
  {Coquidé}}, \bibinfo {author} {\bibfnamefont {L.}~\bibnamefont {Ermann}},
  \bibinfo {author} {\bibfnamefont {J.}~\bibnamefont {Lages}},\ and\ \bibinfo
  {author} {\bibfnamefont {D.}~\bibnamefont {Shepelyansky}},\ }\bibfield
  {title} {\bibinfo {title} {{Influence of petroleum and gas trade on EU
  economies from the reduced Google matrix analysis of UN COMTRADE data}},\
  }\href {https://doi.org/10.1140/epjb/e2019-100132-6} {\bibfield  {journal}
  {\bibinfo  {journal} {{Eur. Phys. J. B}}\ }\textbf {\bibinfo {volume} {92}},\
  \bibinfo {pages} {71} (\bibinfo {year} {2019})}\BibitemShut {NoStop}%
\bibitem [{\citenamefont {Coquidé}\ \emph {et~al.}(2020)\citenamefont
  {Coquidé}, \citenamefont {Lages},\ and\ \citenamefont
  {Shepelyansky}}]{wtncrisis}%
  \BibitemOpen
  \bibfield  {author} {\bibinfo {author} {\bibfnamefont {C.}~\bibnamefont
  {Coquidé}}, \bibinfo {author} {\bibfnamefont {J.}~\bibnamefont {Lages}},\
  and\ \bibinfo {author} {\bibfnamefont {D.}~\bibnamefont {Shepelyansky}},\
  }\bibfield  {title} {\bibinfo {title} {{Crisis contagion in the world trade
  network}},\ }\href {https://doi.org/10.1007/s41109-020-00304-z} {\bibfield
  {journal} {\bibinfo  {journal} {{Appl Netw Sci}}\ }\textbf {\bibinfo {volume}
  {5}},\ \bibinfo {pages} {67} (\bibinfo {year} {2020})}\BibitemShut {NoStop}%
\bibitem [{\citenamefont {Krapivsky}\ \emph {et~al.}(2010)\citenamefont
  {Krapivsky}, \citenamefont {Redner},\ and\ \citenamefont
  {Ben-Naim}}]{krapivsky10}%
  \BibitemOpen
  \bibfield  {author} {\bibinfo {author} {\bibfnamefont {P.~L.}\ \bibnamefont
  {Krapivsky}}, \bibinfo {author} {\bibfnamefont {S.}~\bibnamefont {Redner}},\
  and\ \bibinfo {author} {\bibfnamefont {E.}~\bibnamefont {Ben-Naim}},\ }\href
  {https://doi.org/10.1017/CBO9780511780516} {\emph {\bibinfo {title} {A
  Kinetic View of Statistical Physics}}}\ (\bibinfo  {publisher} {Cambridge
  University Press},\ \bibinfo {year} {2010})\BibitemShut {NoStop}%
\bibitem [{\citenamefont {Blondel}\ \emph {et~al.}(2008)\citenamefont
  {Blondel}, \citenamefont {Guillaume}, \citenamefont {Lambiotte},\ and\
  \citenamefont {Lefebvre}}]{blondel08}%
  \BibitemOpen
  \bibfield  {author} {\bibinfo {author} {\bibfnamefont {V.~D.}\ \bibnamefont
  {Blondel}}, \bibinfo {author} {\bibfnamefont {J.-L.}\ \bibnamefont
  {Guillaume}}, \bibinfo {author} {\bibfnamefont {R.}~\bibnamefont
  {Lambiotte}},\ and\ \bibinfo {author} {\bibfnamefont {E.}~\bibnamefont
  {Lefebvre}},\ }\bibfield  {title} {\bibinfo {title} {Fast unfolding of
  communities in large networks},\ }\href
  {https://doi.org/10.1088/1742-5468/2008/10/P10008} {\bibfield  {journal}
  {\bibinfo  {journal} {Journal of Statistical Mechanics: Theory and
  Experiment}\ }\textbf {\bibinfo {volume} {2008}},\ \bibinfo {pages} {P10008}
  (\bibinfo {year} {2008})},\ \bibinfo {note} {publisher: IOP
  Publishing}\BibitemShut {NoStop}%
\bibitem [{\citenamefont {Dugué}\ and\ \citenamefont {Perez}(2015)}]{dugue15}%
  \BibitemOpen
  \bibfield  {author} {\bibinfo {author} {\bibfnamefont {N.}~\bibnamefont
  {Dugué}}\ and\ \bibinfo {author} {\bibfnamefont {A.}~\bibnamefont {Perez}},\
  }\href {https://hal.archives-ouvertes.fr/hal-01231784} {\emph {\bibinfo
  {title} {Directed {Louvain} : maximizing modularity in directed networks}}},\
  \bibinfo {type} {Research {Report}}\ (\bibinfo  {institution} {Université
  d'Orléans},\ \bibinfo {year} {2015})\BibitemShut {NoStop}%
\bibitem [{\citenamefont {Spivak}()}]{carnegie}%
  \BibitemOpen
  \bibfield  {author} {\bibinfo {author} {\bibfnamefont {V.}~\bibnamefont
  {Spivak}},\ }\href {https://carnegiemoscow.org/commentary/85069} {\bibinfo
  {title} {{Can the Yuan Ever Replace the Dollar for Russia?}}},\ \bibinfo
  {note} {{Carnegie Endowment for International Peace}}\BibitemShut {NoStop}%
\bibitem [{\citenamefont {Kimani}()}]{oilprice}%
  \BibitemOpen
  \bibfield  {author} {\bibinfo {author} {\bibfnamefont {A.}~\bibnamefont
  {Kimani}},\ }\href
  {https://oilprice.com/Energy/Energy-General/China-Looks-To-Expand-Use-Of-Yuan-In-Energy-Deals.html}
  {\bibinfo {title} {{China Looks To Expand Use Of Yuan In Energy Deals}}},\
  \bibinfo {note} {{OilPrice.com}}\BibitemShut {NoStop}%
\bibitem [{\citenamefont {Rosen}()}]{bloomberg2}%
  \BibitemOpen
  \bibfield  {author} {\bibinfo {author} {\bibfnamefont {P.}~\bibnamefont
  {Rosen}},\ }\href
  {https://www.businessinsider.com/china-russian-oil-yuan-steep-discount-price-cap-europe-sanctions-2022-12?r=US&IR=T&utm_source=copy-link&utm_medium=referral&utm_content=topbar}
  {\bibinfo {title} {{China is buying Russian oil at a bigger discount using
  yuan as price cap looms, report says}}},\ \bibinfo {note} {{Business
  Insider}}\BibitemShut {NoStop}%
\bibitem [{\citenamefont {Liu}\ \emph {et~al.}(2019)\citenamefont {Liu},
  \citenamefont {Wang},\ and\ \citenamefont {Woo}}]{liu19}%
  \BibitemOpen
  \bibfield  {author} {\bibinfo {author} {\bibfnamefont {T.}~\bibnamefont
  {Liu}}, \bibinfo {author} {\bibfnamefont {X.}~\bibnamefont {Wang}},\ and\
  \bibinfo {author} {\bibfnamefont {W.~T.}\ \bibnamefont {Woo}},\ }\bibfield
  {title} {\bibinfo {title} {The road to currency internationalization: Global
  perspectives and chinese experience},\ }\href
  {https://doi.org/https://doi.org/10.1016/j.ememar.2018.11.003} {\bibfield
  {journal} {\bibinfo  {journal} {Emerging Markets Review}\ }\textbf {\bibinfo
  {volume} {38}},\ \bibinfo {pages} {73} (\bibinfo {year} {2019})}\BibitemShut
  {NoStop}%
\end{thebibliography}%
\clearpage
\widetext

\textbf{\LARGE Supplementary Information}\\

\textbf{Dollar-Yuan Battle in the World Trade Network}

by Célestin Coquidé, José Lages and Dima L. Shepelyansky

\label{part:suppmat}
%%%%%%%%%% Merge with supplemental materials %%%%%%%%%%
%%%%%%%%%% Prefix a "S" to all equations, figures, tables and reset the counter %%%%%%%%%%

\makeatletter
\renewcommand{\theequation}{S\arabic{equation}}
\renewcommand{\thefigure}{S\arabic{figure}}
\renewcommand{\thetable}{S\arabic{table}}
\setcounter{equation}{0}
\setcounter{figure}{0}
\setcounter{table}{0}
\setcounter{page}{1}
\makeatother

\vfill
\begin{figure}[h]
	\centering
	\includegraphics[width=\textwidth]{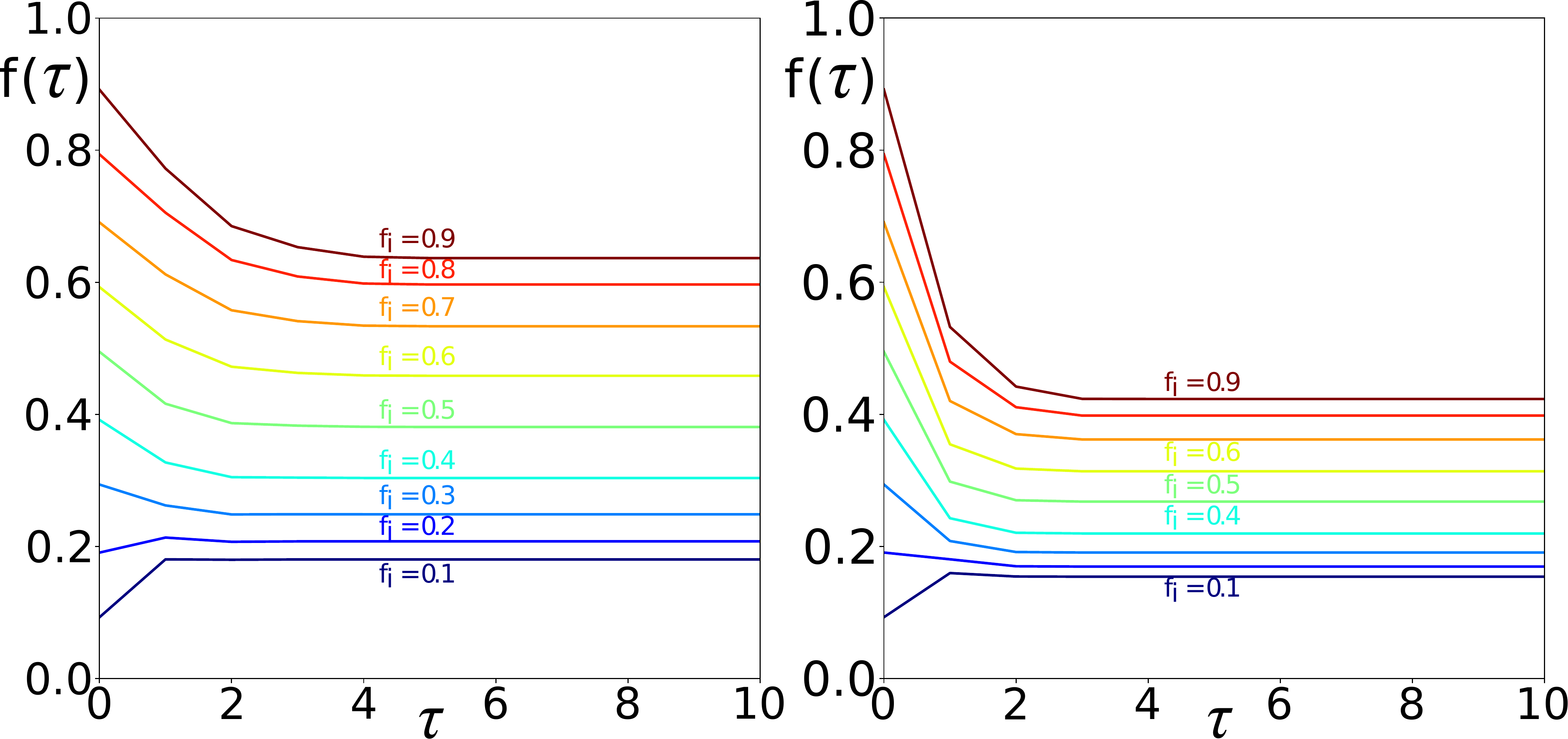}
	\caption{\label{figS1}Evolution of the averaged fraction $f(\tau)$ of countries with USD trade preference as a function of the number of Monte Carlo shake steps $\tau$ for the years 2010 (left panel) and 2019 (right panel).
		The average is performed over $N_r=10^4$ spin configurations.
		The colors indicate the initial fraction $f_i$ of countries with USD trade preference at $\tau=0$.}
\end{figure}
\vfill
\clearpage
\phantom{text}
\vfill
\begin{figure}
	\centering
	\includegraphics[width=\textwidth]{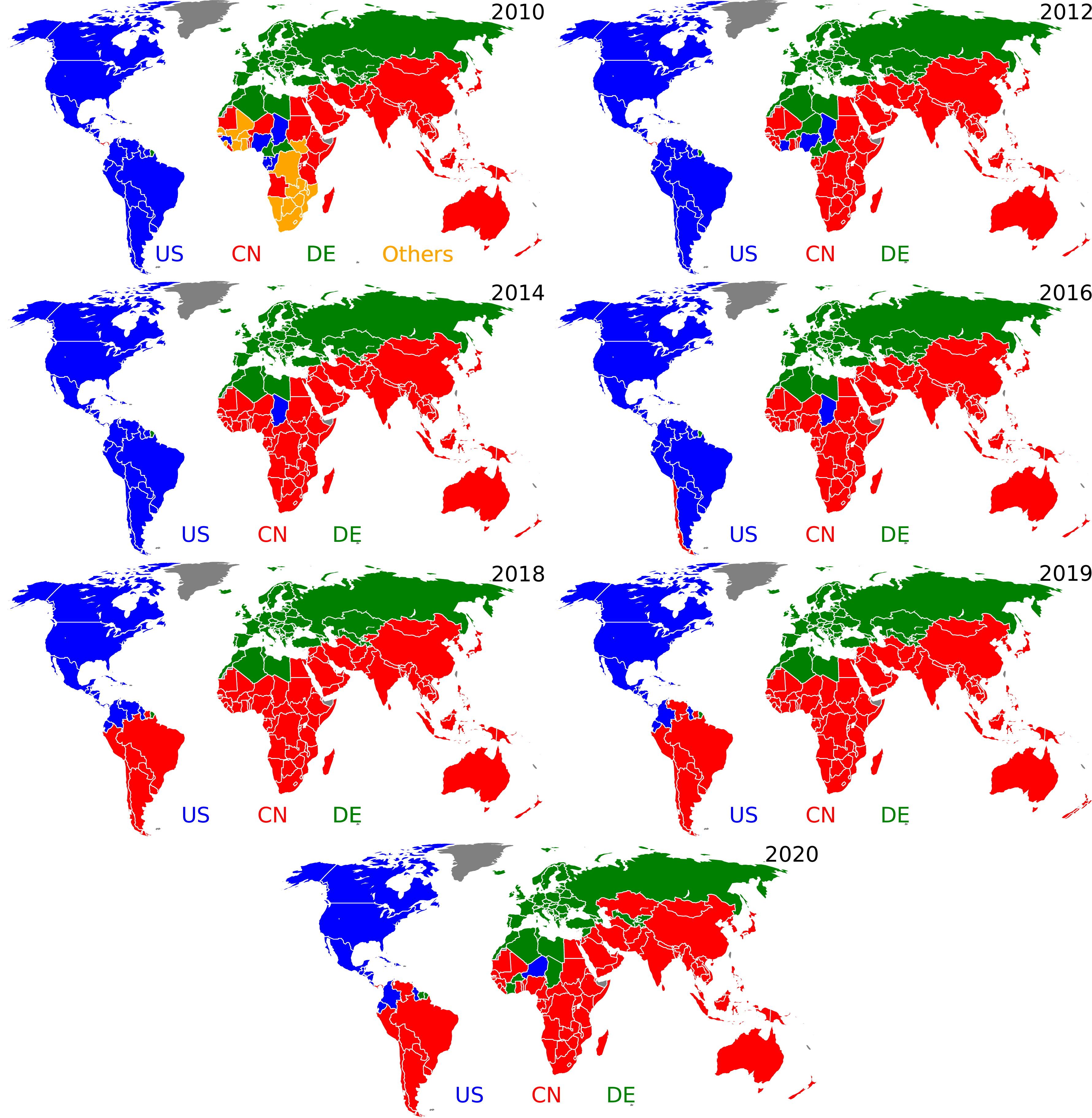}
	\caption{\label{figS2}Geographical distribution of clusters in the WTN for the years 2010, 2012, 2014, 2016, 2018, 2019 and 2020 obtained using the Louvain modularity method \cite{blondel08} with Dugu\'e's algorithm \cite{dugue15}.
		Each cluster is labelled using the ISO2 code of the country with the best import and export trade probabilities $P_c$ and $P^*_c$.
		For all the considered years, the leaders of the clusters are the USA (blue), China (red), and Germany (green).
		The cluster Others (gold) gathers other small size clusters.}
\end{figure}
\vfill
\clearpage
\phantom{text}
\vfill
\begin{figure}
	\centering
	\includegraphics[width=\textwidth]{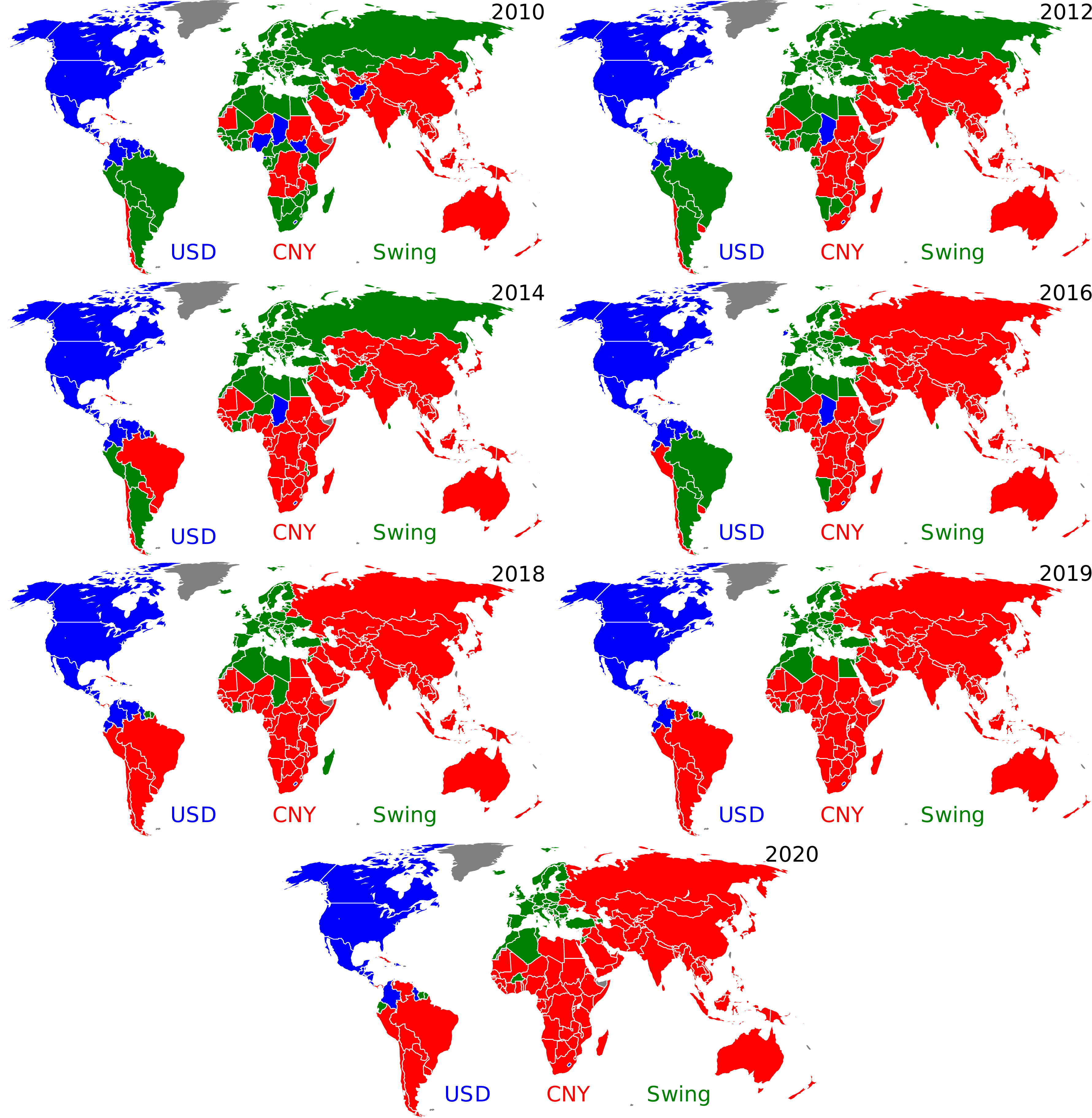}
	\caption{\label{figS3}World map distribution of countries belonging to the USD group
		(blue, hard preference to trade in USD), the CNY group (red, hard preference to trade in CNY)
		and the swing group (green, the TCP can change between USD and CNY depending on the initial conditions). The world maps are shown for the years 2010, 2012, 2014, 2016, 2018, 2019 and 2020.}
\end{figure}
\vfill
\clearpage
\phantom{text}
\vfill
\begin{figure}[t]
	\begin{center}
		\includegraphics[width=\columnwidth]{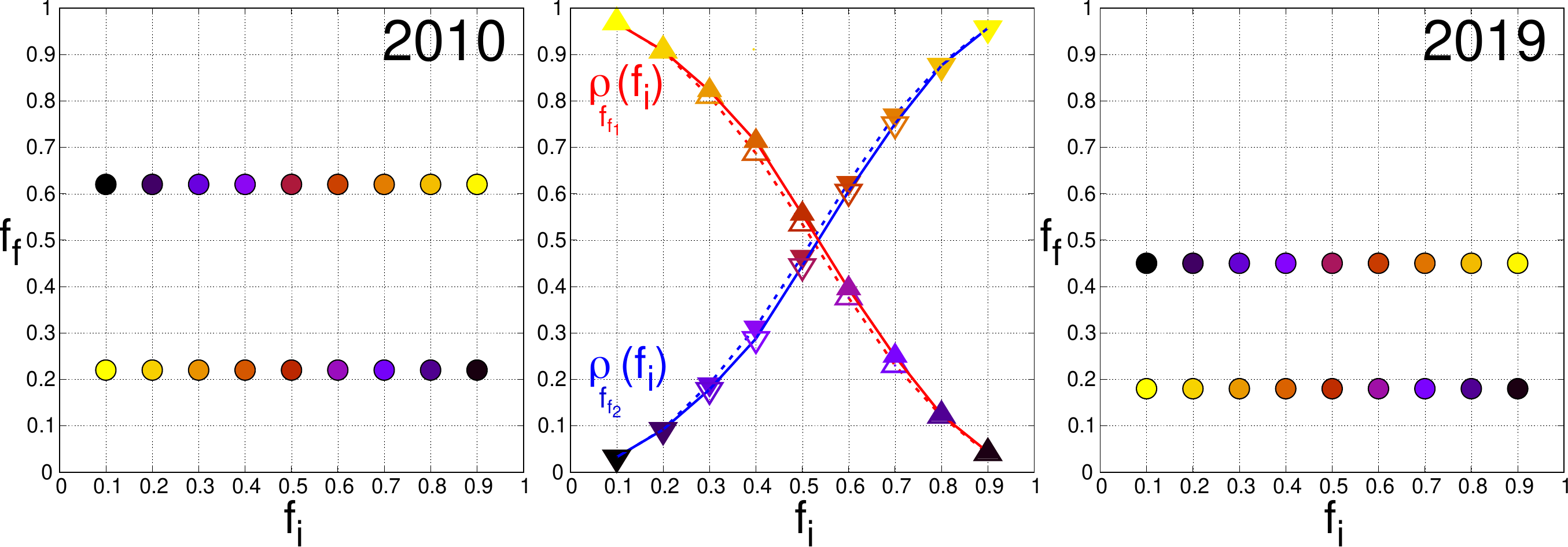}
	\end{center}
	\vglue -0.3cm
	\caption{\label{figS4}Same as Fig.~\ref{fig2} but keeping always USA, Canada, UK, Australia, New Zealand trading in USD and China, Brazil, Russia, India, South Africa in CNY.
		Final fraction $f_f$ of countries preferring to trade in USD as a function of the initial fraction $f_i$ for the years 2010 (left panel) and 2019 (right panel). Left and right panels: for any initial fraction $f_i$, the Monte Carlo procedure converges toward one of two final fractions $f_f$. These two final fractions are $f_{f_1} = 0.22$ and $f_{f_2}=0.62$ in 2010 (left panel)
		and $f_{f_1}=0.18$ and $f_{f_2}=0.45$ in 2019 (right panel). The color of a point ($f_i$,$f_f$) indicates the portion $\rho_{f_f}(f_i)$ of the $N_r=10^4$ initial configurations, with a corresponding initial fraction $f_i$, which attain the final state with the corresponding final fraction $f_f$. The color ranges from black for $\rho_{f_f}(f_i)=0$ (all the countries preferring to trade in CNY rather than in USD) to bright yellow for $\rho_{f_f}(f_i)=1$ (all the countries preferring to trade in USD rather than in CNY). Middle panel: Portion $\rho_{f_f}(f_i)$ of the $N_r=10^4$ initial configurations, the fraction $f_i$ of which initially prefers to trade in USD, which attains the final state with the final fraction $f_f$. The red (blue) curve and the up (down) triangles correspond to the lowest (highest) value of the two final fractions $f_f$. The full (empty) symbols correspond to the year 2019 (2010).}
\end{figure}
\vfill
\clearpage
\phantom{text}
\vfill
\begin{figure}[t]
	\begin{center}
		\includegraphics[width=\columnwidth]{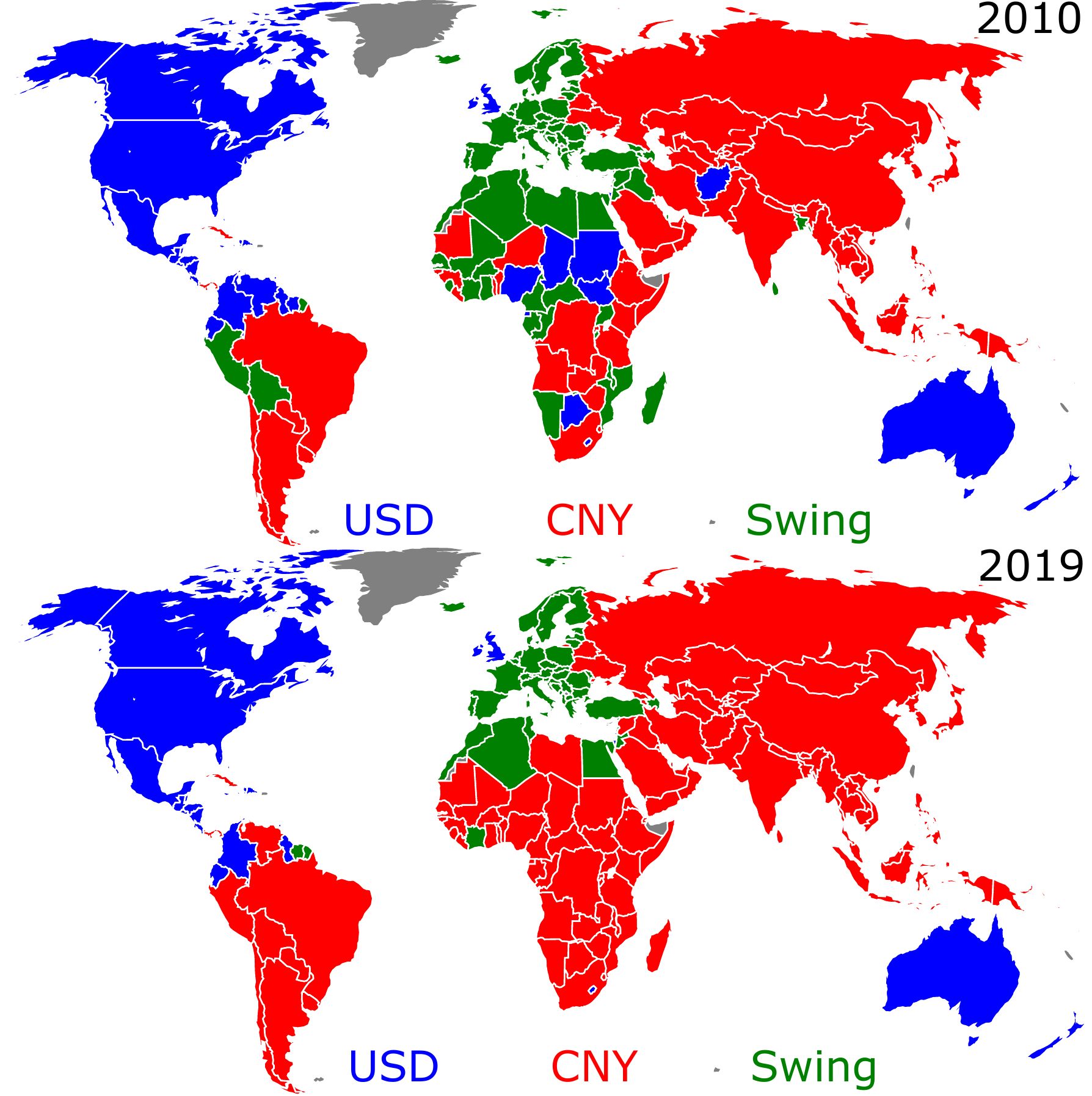}
	\end{center}
	\vglue -0.3cm
	\caption{\label{figS5}
		Same as Fig.~\ref{fig4} but keeping always USA, Canada, UK, Australia, New Zealand trading in USD and China, Brazil, Russia, India, South Africa in CNY.
		World map distribution of countries belonging to the USD group
		(blue, hard preference to trade in USD), the CNY group (red, hard preference to trade in CNY)
		and the swing group (green, the TCP can change between USD and CNY depending on the initial conditions). The world maps are shown for the years 2010 (left panel) and 2019 (right panel).
	}
\end{figure}
\vfill
\clearpage
\phantom{text}
\vfill
\begin{figure}[t]
	\begin{center}
		\includegraphics[width=\columnwidth]{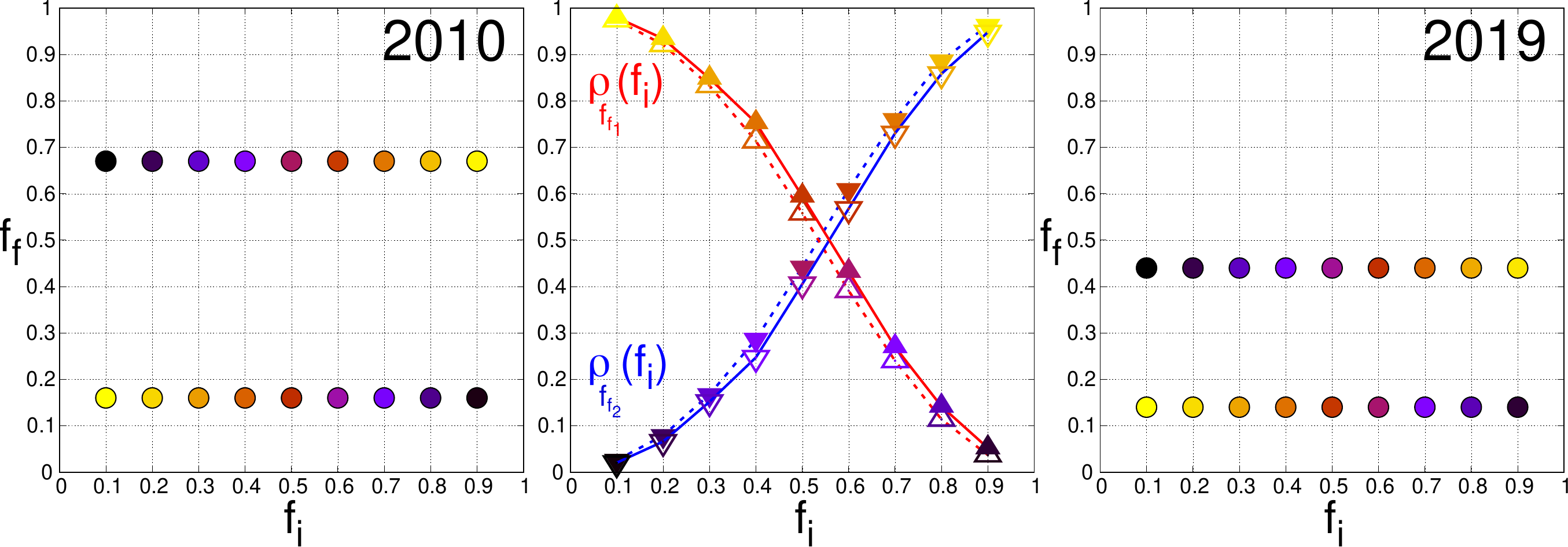}
	\end{center}
	\vglue -0.3cm
	\caption{\label{figS6}Same as Fig.~\ref{fig2} but
		replacing the import trade probabilities $P_c$ and the export trade probabilities $P^*_c$ in
		(1) by the PageRank and the CheiRank probabilities
		obtained from the Google matrix of WTN.
		Final fraction $f_f$ of countries preferring to trade in USD as a function of the initial fraction $f_i$ for the years 2010 (left panel) and 2019 (right panel). Left and right panels: for any initial fraction $f_i$, the Monte Carlo procedure converges toward one of two final fractions $f_f$. These two final fractions are $f_{f_1} = 0.16$ and $f_{f_2}=0.67$ in 2010 (left panel)
		and $f_{f_1}=0.14$ and $f_{f_2}=0.44$ in 2019 (right panel). The color of a point ($f_i$,$f_f$) indicates the portion $\rho_{f_f}(f_i)$ of the $N_r=10^4$ initial configurations, with a corresponding initial fraction $f_i$, which attain the final state with the corresponding final fraction $f_f$. The color ranges from black for $\rho_{f_f}(f_i)=0$ (all the countries preferring to trade in CNY rather than in USD) to bright yellow for $\rho_{f_f}(f_i)=1$ (all the countries preferring to trade in USD rather than in CNY). Middle panel: Portion $\rho_{f_f}(f_i)$ of the $N_r=10^4$ initial configurations, the fraction $f_i$ of which initially prefers to trade in USD, which attains the final state with the final fraction $f_f$. The red (blue) curve and the up (down) triangles correspond to the lowest (highest) value of the two final fractions $f_f$. The full (empty) symbols correspond to the year 2019 (2010).}
\end{figure}
\vfill
\clearpage
\phantom{text}
\vfill
\begin{figure}[t]
	\begin{center}
		\includegraphics[width=\columnwidth]{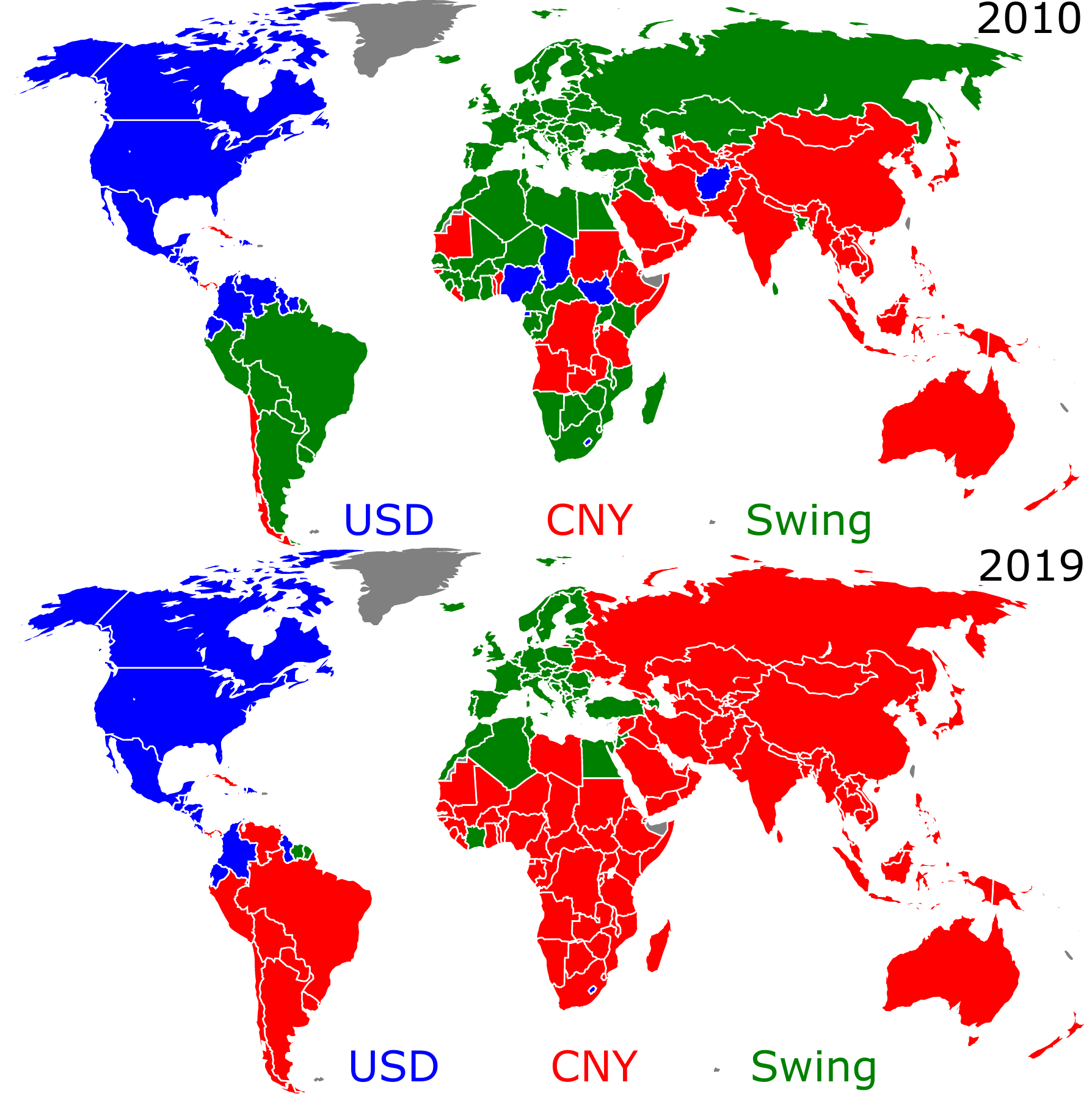}
	\end{center}
	\vglue -0.3cm
	\caption{\label{figS7}
		Same as Fig.~\ref{fig4} but replacing the import trade probabilities $P_c$ and the export trade probabilities $P^*_c$ in
		(1) by the PageRank and the CheiRank probabilities
		obtained from the Google matrix of WTN.
		World map distribution of countries belonging to the USD group
		(blue, hard preference to trade in USD), the CNY group (red, hard preference to trade in CNY)
		and the swing group (green, the TCP can change between USD and CNY depending on the initial conditions). The world maps are shown for the years 2010 (left panel) and 2019 (right panel).
	}
\end{figure}
\clearpage
\begin{table*}
	\centering
	\caption{\label{tab:tabS1}List of the 32 countries belonging to the USD group in 2010. The trade currency preference of these countries is always USD at the end of the relaxation process. The countries are sorted by descending value of $\max\left(P_c,P^*_c\right)$, i.e. the maximum value between the relative import volume $P_c$ and the relative export volume $P_c^*$, and, in case of tie, by descending value of $P^*_c$. The red (green) colored countries switch to the CNY (swing) group in 2019. The countries are represented by their ISO2 codes.}
	
	\begin{tabular}{cc|cc|cc|cc}
		\toprule
		\multicolumn{8}{c}{Countries of the USD group in 2010}\\
		\midrule
		1. & {{US}} & 9. & {GT} & 17. & {JM} & 25. & \textcolor{red}{GQ}\\
		2. & {MX} & 10. & {DO} & 18. & {GY} & 26. & \textcolor{red}{TD}\\
		3. & {CA} & 11. & {HN} & 19. & {BS} & 27. & \textcolor{red}{SS}\\
		4. & \textcolor{green}{IL} & 12. & {SV} & 20. & {HT} & 28. & {LC}\\
		5. & \textcolor{red}{NG} & 13. & {NI} & 21. & \textcolor{green}{SR} & 29. & {KN}\\
		6. & {CO} & 14. & \textcolor{red}{VE} & 22. & {BB} & 30. & {DM}\\
		7. & {EC} & 15. & {TT} & 23. & {LS} & 31. & {GD}\\
		8. & {CR} & 16. & \textcolor{red}{AF} & 24. & {BZ} & 32. & \textcolor{red}{FM}\\
		\bottomrule
	\end{tabular}
\end{table*}
\clearpage
\begin{table*}
	\centering
	\caption{\label{tab:tabS2}List of the 66 countries belonging to the CNY group in 2010. The trade currency preference of these countries is always CNY at the end of the relaxation process. The countries are sorted by descending value of $\max\left(P_c,P^*_c\right)$, i.e. the maximum value between the relative import volume $P_c$ and the relative export volume $P_c^*$, and, in case of tie, by descending value of $P^*_c$. The blue colored countries switch to the USD group in 2019. The countries are represented by their ISO2 codes.}
	
	\begin{tabular}{cc|cc|cc|cc}
		\toprule
		\multicolumn{8}{c}{Countries of the CNY group in 2010}\\
		\midrule
		1. & {{CN}} & 18. & {PK} & 35. & {MR} & 52. & {MV}\\
		2. & {JP} & 19. & {OM} & 36. & {KG} & 53. & {BT}\\
		3. & {KR} & 20. & {KH} & 37. & {PG} & 54. & {GM}\\
		4. & {SG} & 21. & {MM} & 38. & {BJ} & 55. & {DJ}\\
		5. & {IN} & 22. & {UZ} & 39. & {CU} & 56. & {SB}\\
		6. & {VN} & 23. & {BH} & 40. & {YE} & 57. & {TL}\\
		7. & {AE} & 24. & {CD} & 41. & {LR} & 58. & {ER}\\
		8. & {MY} & 25. & {AO} & 42. & {RW} & 59. & {GW}\\
		9. & {TH} & 26. & {ZM} & 43. & {TM} & 60. & {VU}\\
		10. & {AU} & 27. & {PA} & 44. & {MH} & 61. & \textcolor{blue}{WS}\\
		11. & {SA} & 28. & {LA} & 45. & {NP} & 62. & {KI}\\
		12. & {ID} & 29. & {TZ} & 46. & {TJ} & 63. & {PW}\\
		13. & {PH} & 30. & {SD} & 47. & {NE} & 64. & {TV}\\
		14. & {CL} & 31. & {MN} & 48. & {SO} & 65. & {NR}\\
		15. & {NZ} & 32. & {BN} & 49. & {KP} & 66. & \textcolor{blue}{TO}\\
		16. & {KW} & 33. & {ET} & 50. & \textcolor{blue}{CW} & & \\
		17. & {IR} & 34. & {TG} & 51. & \textcolor{blue}{AG} & & \\
		\bottomrule
	\end{tabular}
\end{table*}
\clearpage
\begin{table*}
	\centering
	\caption{\label{tab:tabS3}List of the 96 countries belonging to the swing group in 2010. The trade currency preference of these countries is for all of them either CNY or USD at the end of the relaxation process. The countries are sorted by descending value of $\max\left(P_c,P^*_c\right)$, i.e. the maximum value between the relative import volume $P_c$ and the relative export volume $P_c^*$, and, in case of tie, by descending value of $P^*_c$. The red (blue) colored countries switch to the CNY (USD) group in 2019. The countries are represented by their ISO2 codes.}
	
	\begin{tabular}{cc|cc|cc|cc}
		\toprule
		\multicolumn{8}{c}{Countries of the swing group in 2010}\\
		\midrule
		1. & DE & 25. & \textcolor{red}{UA} & 49. & \textcolor{red}{LY} & 73. & \textcolor{red}{BF}\\
		2. & FR & 26. & \textcolor{red}{AR} & 50. & JO & 74. & \textcolor{red}{MG}\\
		3. & NL & 27. & \textcolor{red}{IQ} & 51. & \textcolor{red}{PY} & 75. & \textcolor{red}{GN}\\
		4. & IT & 28. & \textcolor{red}{BD} & 52. & CI & 76. & \textcolor{red}{ML}\\
		5. & UK & 29. & \textcolor{red}{KZ} & 53. & \textcolor{red}{UY} & 77. & \textcolor{red}{UG}\\
		6. & BE & 30. & SI & 54. & BA & 78. & \textcolor{red}{CG}\\
		7. & CH & 31. & GR & 55. & \textcolor{red}{MZ} & 79. & \textcolor{red}{MW}\\
		8. & ES & 32. & \textcolor{red}{PE} & 56. & MK & 80. & PS\\
		9. & \textcolor{red}{RU} & 33. & EG & 57. & \textcolor{red}{KE} & 81. & FJ\\
		10. & PL & 34. & BG & 58. & \textcolor{red}{BO} & 82. & \textcolor{red}{GA}\\
		11. & \textcolor{red}{BR} & 35. & DZ & 59. & \textcolor{red}{NA} & 83. & \textcolor{red}{SZ}\\
		12. & CZ & 36. & \textcolor{red}{QA} & 60. & IS & 84. & SC\\
		13. & TR & 37. & MA & 61. & MT & 85. & \textcolor{red}{SY}\\
		14. & AT & 38. & LT & 62. & CY & 86. & ME\\
		15. & SE & 39. & \textcolor{red}{BY} & 63. & \textcolor{red}{SN} & 87. & \textcolor{red}{SL}\\
		16. & HU & 40. & RS & 64. & \textcolor{red}{BW} & 88. & \textcolor{red}{BI}\\
		17. & \textcolor{red}{ZA} & 41. & HR & 65. & \textcolor{red}{GE} & 89. & \textcolor{blue}{VC}\\
		18. & IE & 42. & LU & 66. & LB & 90. & AD\\
		19. & DK & 43. & EE & 67. & AL & 91. & CV\\
		20. & NO & 44. & TN & 68. & \textcolor{red}{AM} & 92. & \textcolor{red}{CF}\\
		21. & SK & 45. & LV & 69. & MD & 93. & \textcolor{red}{KM}\\
		22. & RO & 46. & \textcolor{red}{GH} & 70. & \textcolor{red}{CM} & 94. & SM\\
		23. & FI & 47. & AZ & 71. & MU & 95. & ST\\
		24. & PT & 48. & \textcolor{red}{LK} & 72. & \textcolor{red}{ZW} & 96. & PN\\
		\bottomrule
	\end{tabular}
\end{table*}
\clearpage
\begin{table*}
	\centering
	\caption{\label{tab:tabS4}List of the 28 countries belonging to the USD group in 2019. The trade currency preference of these countries is always USD at the end of the relaxation process. The countries are sorted by descending value of $\max\left(P_c,P^*_c\right)$, i.e. the maximum value between the relative import volume $P_c$ and the relative export volume $P_c^*$, and, in case of tie, by descending value of $P^*_c$.	The red (green) colored countries belonged to the CNY (swing) group in 2010.
		The countries are represented by their ISO2 codes.}
	
	\begin{tabular}{cc|cc|cc|cc}
		\toprule
		\multicolumn{8}{c}{Countries of the USD group in 2019}\\
		\midrule
		1.&US&8.&DO&15.&BS&22.&LC\\
		2.&MX&9.&HN&16.&HT&23.&\textcolor{green}{VC}\\
		3.&CA&10.&SV&17.&BB&24.&KN\\
		4.&CO&11.&NI&18.&LS&25.&\textcolor{red}{WS}\\
		5.&EC&12.&TT&19.&BZ&26.&DM\\
		6.&CR&13.&JM&20.&\textcolor{red}{CW}&27.&GD\\
		7.&GT&14.&GY&21.&\textcolor{red}{AG}&28.&\textcolor{red}{TO}\\		
		\bottomrule
	\end{tabular}
\end{table*}
\clearpage
\begin{table*}
	\centering
	\caption{\label{tab:tabS5}List of the 109 countries belonging to the CNY group in 2019. The trade currency preference of these countries is always CNY at the end of the relaxation process. The countries are sorted by descending value of $\max\left(P_c,P^*_c\right)$, i.e. the maximum value between the relative import volume $P_c$ and the relative export volume $P_c^*$, and, in case of tie, by descending value of $P^*_c$.
		The blue (green) colored countries belonged to the USD (swing) group in 2010.
		The countries are represented by their ISO2 codes.
	}
	
	\begin{tabular}{cc|cc|cc|cc}
		\toprule
		\multicolumn{8}{c}{Countries of the CNY group in 2019}\\
		\midrule
		1.&CN&29.&IR&57.&BN&85.&\textcolor{green}{SY}\\
		2.&JP&30.&PK&58.&\textcolor{green}{CM}&86.&NE\\
		3.&KR&31.&OM&59.&\textcolor{blue}{VE}&87.&SO\\
		4.&SG&32.&KH&60.&\textcolor{green}{ZW}&88.&\textcolor{green}{SL}\\
		5.&IN&33.&MM&61.&ET&89.&KP\\
		6.&VN&34.&\textcolor{green}{GH}&62.&\textcolor{green}{BF}&90.&\textcolor{blue}{GQ}\\
		7.&\textcolor{green}{RU}&35.&UZ&63.&TG&91.&MV\\
		8.&AE&36.&\textcolor{green}{LK}&64.&MR&92.&\textcolor{blue}{TD}\\
		9.&MY&37.&\textcolor{green}{LY}&65.&\textcolor{blue}{AF}&93.&\textcolor{green}{BI}\\
		10.&TH&38.&BH&66.&KG&94.&BT\\
		11.&AU&39.&\textcolor{green}{PY}&67.&PG&95.&\textcolor{blue}{SS}\\
		12.&\textcolor{green}{BR}&40.&\textcolor{green}{UY}&68.&\textcolor{green}{MG}&96.&GM\\
		13.&SA&41.&\textcolor{green}{MZ}&69.&BJ&97.&DJ\\
		14.&ID&42.&\textcolor{green}{KE}&70.&\textcolor{green}{GN}&98.&SB\\
		15.&\textcolor{green}{ZA}&43.&CD&71.&\textcolor{green}{ML}&99.&TL\\
		16.&PH&44.&\textcolor{green}{BO}&72.&CU&100.&ER\\
		17.&CL&45.&AO&73.&YE&101.&\textcolor{green}{CF}\\
		18.&\textcolor{blue}{NG}&46.&ZM&74.&\textcolor{green}{UG}&102.&GW\\
		19.&\textcolor{green}{UA}&47.&\textcolor{green}{NA}&75.&LR&103.&\textcolor{green}{KM}\\
		20.&\textcolor{green}{AR}&48.&PA&76.&RW&104.&VU\\
		21.&\textcolor{green}{IQ}&49.&LA&77.&\textcolor{green}{CG}&105.&\textcolor{blue}{FM}\\
		22.&\textcolor{green}{BD}&50.&\textcolor{green}{SN}&78.&TM&106.&KI\\
		23.&\textcolor{green}{KZ}&51.&TZ&79.&\textcolor{green}{MW}&107.&PW\\
		24.&NZ&52.&\textcolor{green}{BW}&80.&MH&108.&TV\\
		25.&\textcolor{green}{PE}&53.&\textcolor{green}{GE}&81.&NP&109.&NR\\
		26.&KW&54.&SD&82.&\textcolor{green}{GA}&&\\
		27.&\textcolor{green}{QA}&55.&MN&83.&TJ&&\\
		28.&\textcolor{green}{BY}&56.&\textcolor{green}{AM}&84.&\textcolor{green}{SZ}&&\\
		\bottomrule
	\end{tabular}
\end{table*}
\clearpage
\begin{table*}
	\centering
	\caption{\label{tab:tabS6}List of the 57 countries belonging to the swing group in 2019. The trade currency preference of these countries is for all of them either CNY or USD at the end of the relaxation process. The countries are sorted by descending value of $\max\left(P_c,P^*_c\right)$, i.e. the maximum value between the relative import volume $P_c$ and the relative export volume $P_c^*$, and, in case of tie, by descending value of $P^*_c$.
		The blue colored countries belonged to the USD group in 2010.
		The countries are represented by their ISO2 codes.}
	
	\begin{tabular}{cc|cc|cc|cc}
		\toprule
		\multicolumn{8}{c}{Countries of the swing group in 2019}\\
		\midrule
		1.&DE&16.&DK&31.&HR&46.&MD\\
		2.&FR&17.&NO&32.&LU&47.&MU\\
		3.&NL&18.&SK&33.&EE&48.&PS\\
		4.&IT&19.&RO&34.&TN&49.&FJ\\
		5.&UK&20.&FI&35.&LV&50.&\textcolor{blue}{SR}\\
		6.&BE&21.&PT&36.&AZ&51.&SC\\
		7.&CH&22.&\textcolor{blue}{IL}&37.&JO&52.&ME\\
		8.&ES&23.&SI&38.&CI&53.&AD\\
		9.&PL&24.&GR&39.&BA&54.&CV\\
		10.&CZ&25.&EG&40.&MK&55.&SM\\
		11.&TR&26.&BG&41.&IS&56.&ST\\
		12.&AT&27.&DZ&42.&MT&57.&PN\\
		13.&SE&28.&MA&43.&CY&&\\
		14.&HU&29.&LT&44.&LB&&\\
		15.&IE&30.&RS&45.&AL&&\\
		\bottomrule
	\end{tabular}
\end{table*}

\end{document}